\documentclass[letterpaper,twocolumn,10pt]{article}
\usepackage{usenix-2020-09}

\usepackage{amsmath,amsfonts}
\usepackage{graphicx}
\usepackage{subcaption}
\usepackage{tabularx}
\usepackage{multirow}
\usepackage{soul}
\usepackage[english]{babel}
\usepackage{blindtext}

\usepackage{authblk}
\makeatletter
\renewcommand\AB@affilsepx{, \protect\Affilfont}
\makeatother

\usepackage[compact]{titlesec}
\titlespacing*{\section}
{0pt}{3pt}{3pt}
\titleformat*{\subsection}{\bfseries}
\titlespacing*{\subsection}
{0pt}{3pt}{3pt}

\titlespacing*{\subsubsection}
{0pt}{3pt}{3pt}

\begin{document}
\date{}
%-------------------------------------------------------------------------------

%don't want date printed

\title{\Large \bf NN-Defined Modulator: Reconfigurable and Portable Software Modulator on IoT Gateways}

%\title{\vspace{-20pt}\Large \bf mmComb: High-speed mmWave Commodity WiFi Backscatter}

%\title{\Large \bf NN-Defined Modulator: Reconfigurable and Portable Software Modulator on IoT Gateways}

%for single author (just remove % characters)
\author[1]{\rm Jiazhao Wang}
\author[1]{\rm Wenchao Jiang}
\affil[1]{\it Singapore University of Technology and Design}

\author[2]{\rm Ruofeng Liu}
\affil[2]{\it University of Minnesota}

\author[3]{\rm Bin Hu}
\affil[3]{\it University of Southern California}

\author[4]{\rm Demin Gao}
\affil[4]{\it Nanjing Forestry University}

\author[5]{\rm Shuai Wang}
\affil[5]{\it Southeast University}

% \author[\textdagger]{\rm Ruofeng Liu}
% \author{
% {\rm Jiazhao Wang \quad\quad \rm Wenchao Jiang}\\
% Singapore University of Technology and Design
% \and
% {\rm Ruofeng Liu}\\
% \and
% {\rm Bin Hu}\\
% University of Minnesota
% \and
% {\rm Demin Gao}\\
% Nanjing Forestry University
% \and
% {\rm Shuai Wang}\\
% Southeast University
% }

% \author{
% {\rm Jiazhao Wang}\\
% Singapore University of Technology and Design
% \and
% {\rm Wenchao Jiang}\\
% Singapore University of Technology and Design
% \and
% {\rm Ruofeng Liu}\\
% University of Minnesota
% \and
% {\rm Bin Hu}\\
% University of Minnesota
% \and
% {\rm Demin Gao}\\
% Nanjing Forestry University
% \and
% {\rm Shuai Wang}\\
% Southeast University
% }
% end author

\maketitle

%-------------------------------------------------------------------------------
\begin{abstract}

A physical-layer modulator is a vital component for an IoT gateway to map the symbols to signals. However, due to the soldered hardware chipsets on the gateway's motherboards or the diverse toolkits on different platforms for the software radio, the existing solutions either have limited extensibility or are platform-specific. Such limitation is hard to ignore when modulation schemes and hardware platforms have become extremely diverse. This paper presents a new paradigm of using neural networks as an abstraction layer for physical layer modulators in IoT gateway devices, referred to as NN-defined modulators. Our approach addresses the challenges of extensibility and portability for multiple technologies on various hardware platforms. The proposed NN-defined modulator uses a model-driven methodology rooted in solid mathematical foundations while having native support for hardware acceleration and portability to heterogeneous platforms. We conduct the evaluation of NN-defined modulators on different platforms, including Nvidia Jetson Nano and Raspberry Pi. Evaluations demonstrate that our NN-defined modulator effectively operates as conventional modulators and provides significant efficiency gains (up to $4.7\times$ on Nvidia Jetson Nano and $1.1\times$ on Raspberry Pi), indicating high portability. Furthermore, we show the real-world applications using our NN-defined modulators to generate ZigBee and WiFi packets, which are compliant with commodity TI CC2650 (ZigBee) and Intel AX201 (WiFi NIC), respectively. 

\end{abstract}

\section{Introduction}
\begin{figure}[t]
    % \vspace{-2mm} 
    \centerline{\includegraphics[width=0.9\linewidth]{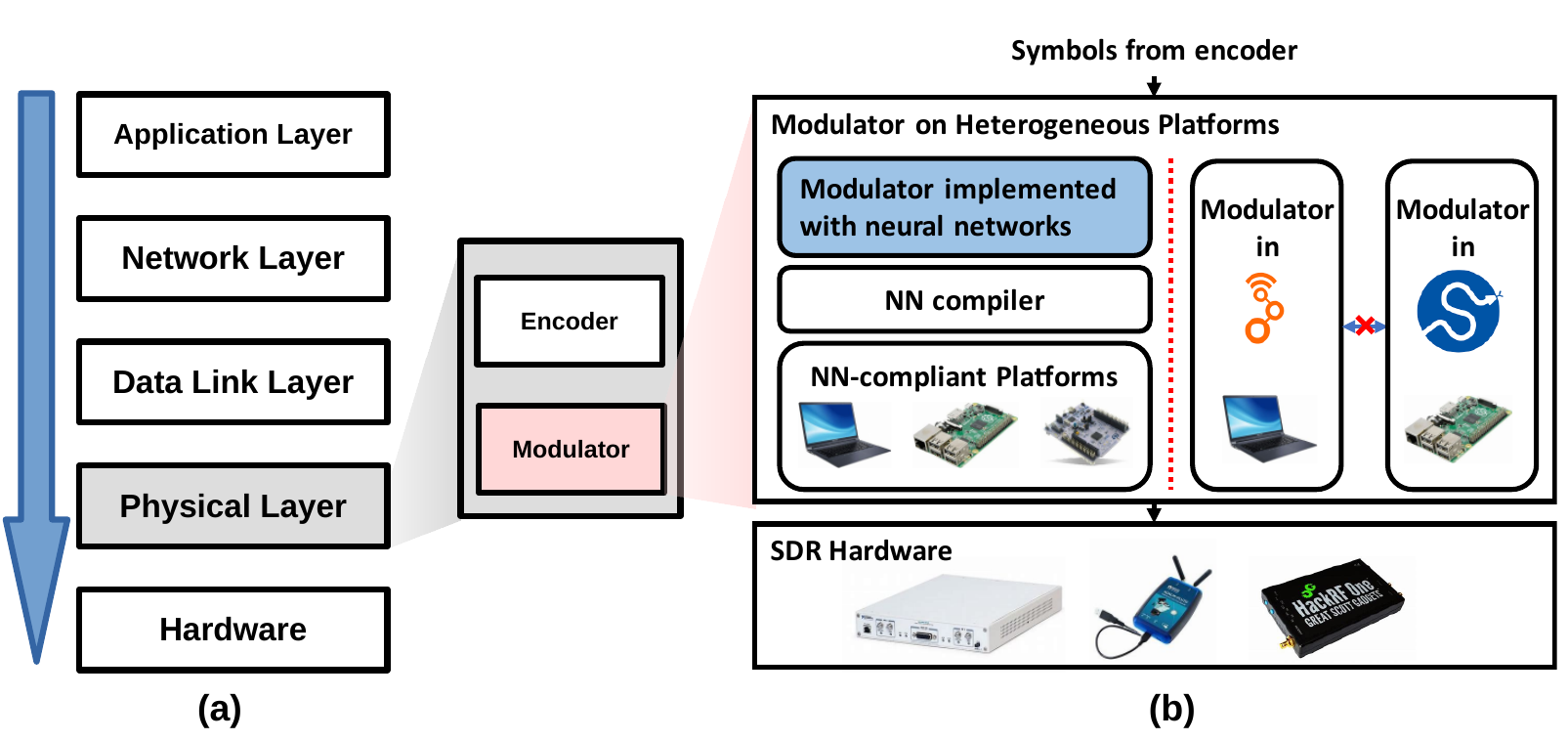}}
    \caption{(a) Simplified layered model for IoT protocol stack. Our design is focused on the Physical layer. (b) Left: deployment of the modulator based on Neural Network. Right: deployment of the modulator based on SDR toolkits.}
    \label{fig:IoTStack}
    \vspace{-5mm} 
\end{figure}

In recent years, we have observed the swift progression of the Internet of Things (IoT), transitioning from theoretical concepts to tangible reality. IoT's objective is to connect many devices, such as sensors and actuators, globally via various Physical (PHY) layer technologies. These technologies are tailored to suit IoT connections based on factors like throughput, power consumption, and coverage area. For instance, IEEE 802.15.4~\cite{ieee802154} is specifically designed for short-range, low-rate IoT connections, while NB-IoT~\cite{NBIoT} is intended for broader, low-power IoT connections. The IoT gateway functions as a central hub, establishing wireless communication links with IoT devices and bridging them to the rest of the Internet. Within PHY of IoT communication (Figure~\ref{fig:IoTStack}a), the modulator plays a vital role in generating signals for data transfer to transmit over the air. 

Therefore, it is essential for the gateway to be flexible, allowing it to support a variety of wireless technologies used by IoT devices and even accommodate emerging technologies for future readiness. However, numerous existing gateway designs~\cite{hardwario, MultiGateway2018, gioia2016amber,morabito2018legiot} employ hardware-based solutions, where wireless technologies are integrated into dedicated chipsets, which are either soldered onto the gateway's motherboard or connected through extension ports. Such hardware-based solutions offer limited adaptability, as their functions are fixed upon manufacturing. Software Defined Radio (SDR) is introduced as a flexible alternative for IoT gateways to address these limitations. Users can implement both current and emerging wireless transceivers as software, surpassing the extensibility of hardware-based solutions.

% \begin{figure}[!htbp]
%      \centering
%      \begin{subfigure}[b]{0.4\linewidth}
%          \centering
%          \includegraphics[width=0.9\linewidth]{Figures/Overview/LayerModel.pdf}
%          \caption{}
%          \label{fig:IoTStack}
%      \end{subfigure}
%      \begin{subfigure}[b]{0.5\linewidth}
%          \centering
%          \includegraphics[width=0.9\linewidth]{Figures/Overview/SystemOverview_2.pdf}
%          \caption{}
%          \label{fig:IoTSDR}
%      \end{subfigure}     
%      \caption{(a) Simplified layered model for IoT communication, (b) Left: deployment of the modulator based on Neural Network. Right: deployment of the modulator based on SDR toolkits.}
%      \vspace{-5mm}
% \end{figure}

Despite the advantage of flexibility, SDR-based gateway design comes with several drawbacks. SDR-based designs consist of the radio frequency (RF) front-end and the computing device serving as the host device for the software radio application. The software radio application requires using signal processing toolkits or libraries, like GNURadio~\cite{GNURadio} and SciPy~\cite{scipy}. Owing to the variety of development tools and host platforms, transferring the same functionality to a new platform demands a considerable learning curve and extensive effort in software development. Meanwhile, software radio comes at the cost of efficiency loss as we shift the signal processing from specialized hardware to general software systems. Many researchers and developers intend to optimize the software radio with the capability of hardware acceleration~\cite{li2015accelerating,kazaz2017hardware,georgis2020dsp}. However, these works are targeted at specific platforms and require a considerable learning curve and extensive effort during development. Given the diversity of platforms and toolkits, deploying a highly efficient software modulator on multiple platforms becomes challenging.

To address these issues, we propose to develop a novel framework that facilitates the design of software transmitters on a variety of IoT gateways using neural networks. This approach would maintain the flexibility needed to accommodate a wide range of transmitters while enhancing portability and efficiency on different platforms. Our work is motivated by several interesting observations: \textbf{i)} the neural network module is widely supported across diverse hardware platforms due to the flourishing AI technologies, and \textbf{ii)} our research demonstrates that signal processing blocks within a modulator can be equivalently implemented using neural network models. These insights have led to the development of our innovative neural network-defined modulator~\footnote{The code for reproduction is available at the anonymous repository: https://github.com/Repo4Sub/NSDI2024}, which offers a flexible and portable design for IoT gateways. The complete architecture of the design is depicted in Figure~\ref{fig:IoTStack}b. The proposed neural network-defined modulator functions are implemented by a unified neural network framework that can take advantage of accelerators across various platforms. In essence, the unified neural network framework operates as an abstraction layer for modulation tasks across heterogeneous platforms.

In summary, the original contributions of this paper are listed as follows:
\begin{itemize}
\item Conceptually, we propose an NN-defined physical layer modulator, which achieves high flexibility and extensibility to support multiple modulation schemes, and portability and efficiency on heterogeneous platforms. 

\item Technically, we adopt a model-driven approach to build the NN-defined modulators. The structure and parameters in the NN-defined modulators are rooted in a solid mathematics foundation from the modulation model.

\item Experimentally, we deploy the NN-defined modulators on multiple hardware platforms (e.g., Nvidia Jetson Nano, Raspberry Pi) with extensive evaluations. We also employ our NN-defined modulators into the workflow of the IoT gateway to generate protocol-compliant signals, including ZigBee and WiFi.
\end{itemize}

% Our results have shown that the NN-defined modulator outperforms the conventional SDR modulators in terms of portability and runs up to $28$ times faster than the conventional modulator under specific settings.

% The rest of this article is organized as follows. In Section~\ref{sec:motivation}, the motivation is given, followed by an overview of the design in Section \ref{sec:overview}. In Section~\ref{sec:template}, the modulation model and its equivalent neural network implementation are reviewed. Section~\ref{sec:modulator} and Section~\ref{sec:learning} discuss construction details for the NN-defined modulators.  Section~\ref{sec:portability} describes the portability of the proposed design. Section~\ref{sec:eval} demonstrates the implementation and some evaluation results. Section~\ref{sec:rela} lists relevant research works. Finally, Section~\ref{sec:concl} gives the discussion on the future work.
\section{Motivation}\label{sec:motivation}
\subsection{Problem Statement}
Modern IoT gateways strive to offer adaptable transmitters to address the ever-evolving landscape of IoT connectivity technologies. Present IoT gateway solutions can be classified into hardware-based gateways and those based on software-defined radio (SDR). Hardware-based solutions, as the term implies, combine numerous chips/modules tailored for various connectivity technologies on a single board~\cite{hardwario, MultiGateway2018, gioia2016amber,morabito2018legiot}. While these hardware chips/modules exhibit merits such as cost-effectiveness and efficiency, they are limited by their lack of adaptability. This limitation stems from the fixed nature of technologies within the chips/modules and the restricted capacity for users to alter connectivity technologies. Gateway platforms can take the form of diverse devices, including personal computers, edge servers, and, increasingly, embedded computers, which can function as host devices for SDR. Consequently, it is possible to develop software radio for distinct IoT technologies, surpassing hardware-based solutions in terms of flexibility. Nevertheless, the multitude of development tools and deployment platforms for software radio can impede portability. For instance, GNU Radio establishes the signal processing blocks required for software radio construction, yet porting GNU Radio to embedded computers for IoT gateways is challenging, necessitating recompilation for target devices~\cite{grcon2021}. Moreover, platform/toolkit-specific implementations often depend on optimized designs that exploit acceleration capabilities, potentially resulting in efficiency loss when transferring SDR-based solutions to new platforms. For example, cuSignal~\cite{CuSignal} is a GPU-accelerated signal processing library, exclusively designed for devices equipped with NVIDIA GPUs, thus not providing a universal solution.

\subsection{Opportunities}
Our design is inspired by the extensive integration of AI frameworks and hardware across diverse computing platforms, which can serve as IoT gateways. Hardware manufacturers continuously enhance their devices to facilitate neural network deployments, incorporating specialized instruction sets~\cite{Upedge} and distinct hardware accelerators~\cite{jetson}, along with programming libraries that capitalize on these features. Concurrently, nearly all mainstream machine learning frameworks, such as Tensorflow~\cite{tensorflow2015} and PyTorch~\cite{pytorch}, endeavor to function across various operating systems and hardware architectures. Additionally, these frameworks encapsulate low-level acceleration libraries, promoting developers to speed up the execution of neural network models.

\begin{figure}[!ht]
\vspace{-2mm}
     \centering
     \begin{subfigure}[b]{0.4\linewidth}
         \centering
         \includegraphics[width=\linewidth]{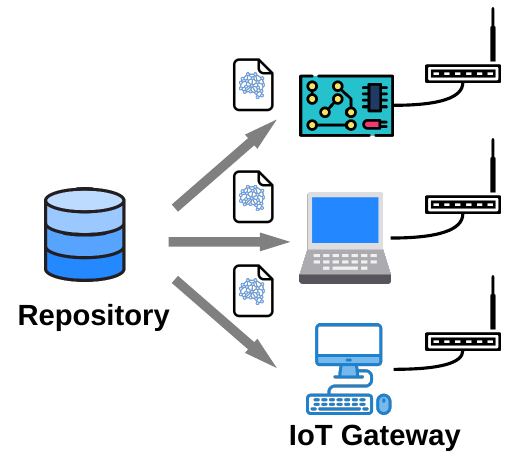}
         \caption{}
         \label{fig:NNeasyDeployment}
     \end{subfigure}
     \begin{subfigure}[b]{0.4\linewidth}
         \centering
         \includegraphics[width=0.8\linewidth]{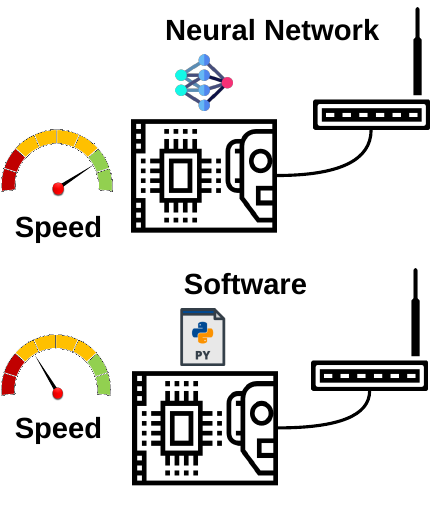}
         \caption{}
         \label{fig:NNhighEfficiency}
     \end{subfigure}
     \vspace{-3mm}
     \caption{(a) Different devices can retrieve various neural network models to update modulation schemes. (b) Neural network modulators can be accelerated to achieve better efficiency compared with software modulators.}
     \vspace{-5mm}
\end{figure}

By constructing transmitters as neural network modules with widespread support, we can attain not only extensibility for an array of technologies but also portability and efficiency on platforms compatible with neural networks. For one thing, a gateway device can always update its supported modulation schemes by retrieving the corresponding neural network implementation from the repository server (Figure~\ref{fig:NNeasyDeployment}). Simultaneously, the neural-network-defined modulators are expected to achieve superior efficiency compared to traditional software modulators (Figure~\ref{fig:NNhighEfficiency}), which are blessed by the advantages of the hardware accelerators.

\subsection{Challenges}
The primary technical challenge involves integrating signal processing blocks into neural network models. One direct method is to utilize general-purpose neural network models, such as fully-connected (FC) layers, as in the literature~\cite{ye2017power, Data:Oshea:17}. Nevertheless, we contend that this approach has two principal disadvantages compared to the traditional digital modulation model. The operational mechanism of a general-purpose machine learning model is often perceived as a black-box approach \cite{he2019model}, which raises concerns about its reliability. 
% Secondly, such a neural network demands an extensive volume of training data to converge, which is typically too complex, bulky, and energy-intensive for IoT gateway applications.

\begin{figure}[!htbp]
    \vspace{-3mm} 
    \centerline{\includegraphics[width=0.6\linewidth, keepaspectratio=true]{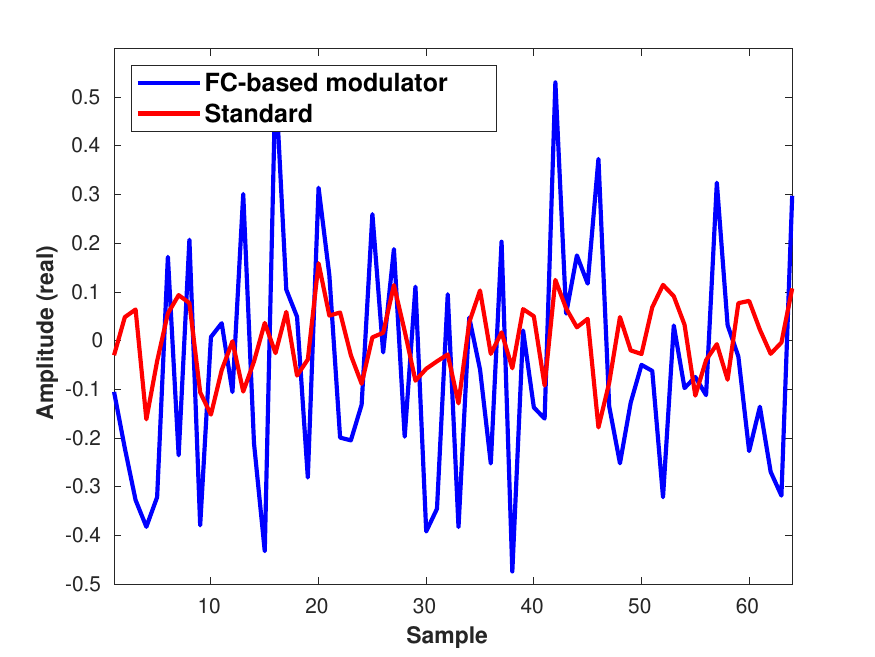}}
    \vspace{-2mm} 
    \caption{Waveform (real part) comparison of FC-based modulator and standard $64$-S.C. OFDM modulator.}
    \label{fig:OFDMWaveform_sample}
    \vspace{-5mm} 
\end{figure}

To illustrate this, we present a straightforward example of modulators based on general-purpose neural networks. We develop an FC-based neural network model to modulate OFDM symbols and train it using the dataset gathered from the standard $64$-S.C. (subcarrier) OFDM modulator. The FC-based OFDM modulator converges to a Mean Squared Error (MSE) loss of $1.5 \times 10^{-6}$ for the training set, signifying that the generated signals from training symbols closely resemble the corresponding training signals. However, it fails to modulate new OFDM symbols from the test set. The produced signal samples are depicted in Figure~\ref{fig:OFDMWaveform_sample}. The output from the FC-based modulator substantially deviates from the standard signals. Although this is a simple case study, we can deduce from these results that the neural network ought to be meticulously designed and executed to achieve modulation tasks.

Our research advocates for a model-driven strategy that integrates domain-specific knowledge of modulation techniques into the neural network design process. Instead of employing general-purpose neural network models or creating tailored neural network layers, our objective is to interpret fundamental neural network layers using domain-specific expertise regarding modulation schemes. By assembling the modulator with these neural network layers, which are comprehensively supported and efficiently implemented across diverse frameworks and platforms, we accomplish an interpretable, lightweight, and efficient neural network-based implementation for software modulators.

\begin{figure}[t]
    \centering
    \includegraphics[width=\linewidth]{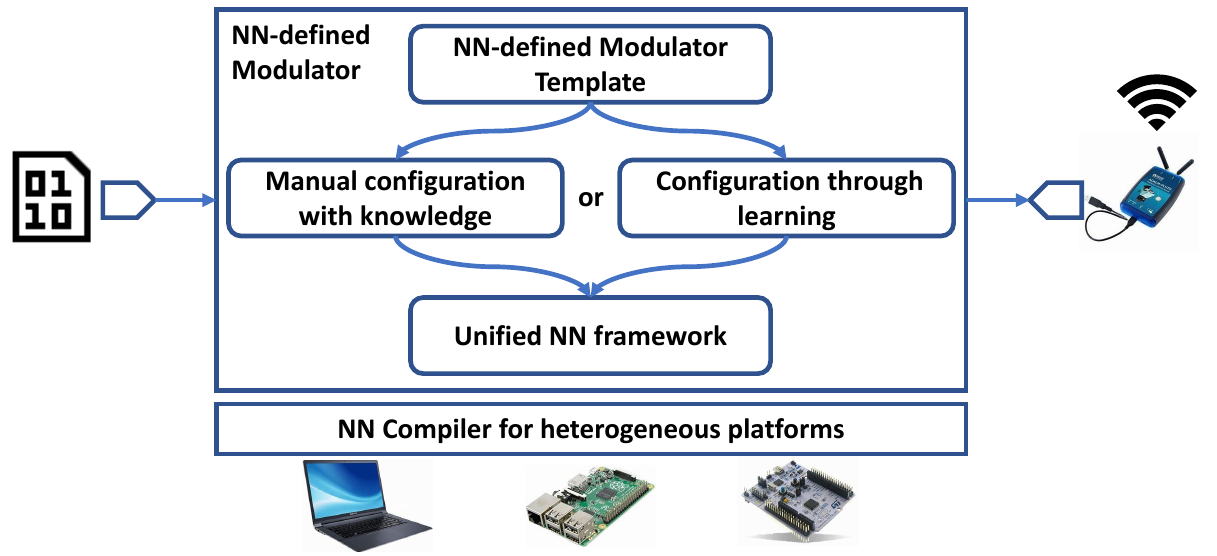}
    \vspace{-3mm} 
    \caption{Architecture of NN-defined modulator.}
    \label{fig:deisgnoverview}
    \vspace{-5mm} 
\end{figure}

The architecture of the proposed Neural-Network-Defined (NN-defined) modulator is depicted in Figure~\ref{fig:deisgnoverview}. As in the figure, a modulator template~(Section~\ref{sec:template}) rooted in solid mathematical foundations can be configured to implement specific modulation schemes either manually as in Section~\ref{sec:modulator}, or in a learning manner as in Section~\ref{sec:learning}. Next, the NN-defined modulator will be transformed into a unified NN framework capable of executing across heterogeneous platforms (Section~\ref{sec:portability}). The unified NN framework can be deployed onto various platforms and incorporated into the transmission pipeline~(Section~\ref{sec:eval}).

\section{Template of NN-defined Modulator}\label{sec:template}
In this section, we discuss how to use a model-driven methodology to construct neural networks for modulation tasks based on the underlying digital modulation models.

\subsection{Mathematical Foundation of Digital Modulation}
In wireless communication, a transmitter uses a modulator to convert symbols to signals before transmitting them to the air. The modulation process is usually analyzed through the Signal Space Analysis~\cite{Proakis2007digital,goldsmith2005wireless}, which is widely adopted in modeling amplitude/phase modulation techniques or named as linear modulation~\cite{goldsmith2005wireless}, including pulse amplitude modulation (PAM), phase-shift keying (PSK), and quadrature amplitude modulation (QAM). And the concepts are also applicable in modeling multicarrier modulation schemes, like OFDM.

Based on this method, a signal $S_i(t)$ modulated from a symbol $s_i$ is considered a linear combination of the set of basis functions. The synthesis process is given as
\begin{equation}
    S_i(t)=\sum_{j=1}^{N}s_{ij}\phi_j(t)
    \label{eq:modulation_cont}
\end{equation}
where $\phi_j(t) \in \{\phi(t)\}^N$ is the $j$-th function in the set of $N$ basis functions and $s_{ij}$ is the $j$-th elements of the $N$-dimensional vector representation of the input symbol $s_i$. The basis functions and format of the symbol can be diverse and determine the different modulation schemes. 

\subsection{Modulator Template via Neural Network}\label{subsec:NNfir}
After modeling the modulation process, we start to fit such a mathematical model into the neural network design to construct the proposed NN-defined modulator template. 

\subsubsection{Discrete-time Modulation Model}  To accommodate the model into the neural networks, we first derive the discrete-time representation of the general model. The $N$-dimensional symbol vector $s_i$ will be modulated to a series of signal samples as: 
\begin{equation}
    S_i[n] = \sum_{j=1}^{N}s_{ij}\phi_j[n]
    \label{eq:modulation_disc}
\end{equation}
where $\phi_j[n]$ is the discrete-time form of the basis functions. The symbols are processed sequentially, and the signal samples are concatenated into the final modulated signals for the whole symbol sequence, given as: 
\begin{equation}
    S[n]=\sum_{i}S_i[n-iL]
    \label{eq:modulation_sps}
\end{equation}
where $L$ is the number of \textit{samples per symbol}, meaning that $L$ samples represent one symbol in the final modulated signals. 

To extend Equation~(\ref{eq:modulation_disc}) to a complex I/Q signal, we have:  
\begin{equation}
\begin{aligned}
    &S_I[n] + jS_Q[n] = \mathrm{Re}\{S_i[n]\} + j\mathrm{Im}\{S_i[n]\}   \\
    &= \sum_{j=1}^{N}[\mathrm{Re}\{s_{ij}\} + j\mathrm{Im}\{s_{ij}\}][\mathrm{Re}\{\phi_j[n]\} + j\mathrm{Im}\{\phi_j[n]\}] \\
    &= \sum_{j=1}^{N}\mathrm{Re}\{s_{ij}\}\mathrm{Re}\{\phi_j[n]\}-\sum_{j=1}^{N}\mathrm{Im}\{s_{ij}\}\mathrm{Im}\{\phi_j[n]\} \\
    &+j(\sum_{j=1}^{N}\mathrm{Re}\{s_{ij}\}\mathrm{Im}\{\phi_j[n]\}+\sum_{j=1}^{N}\mathrm{Im}\{s_{ij}\}\mathrm{Re}\{\phi_j[n]\})
\end{aligned}
\label{eq:modulation_part}
\end{equation}
where $S_I$ is the In-Phase signal and $S_Q$ is the Quadrature. We can observe Equation (\ref{eq:modulation_part}) is composed of multiple $\sum_{j=1}^{N}s_{ij}\phi_j[n]$ patterns. 

\subsubsection{Basics of Transposed Convolutional Layer} 

\begin{figure}[!htbp]
    % \vspace{-2mm} 
    \centering
    \includegraphics[width=0.6\linewidth]{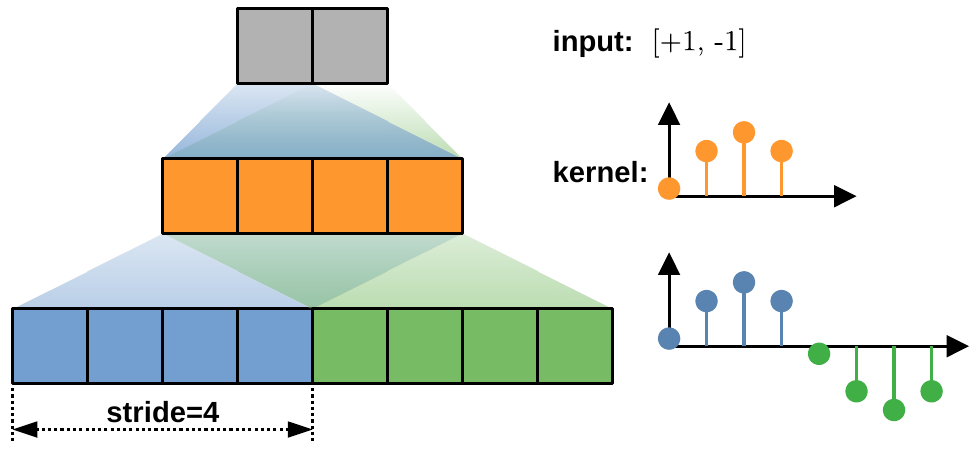}
    \caption{Diagram of the basic operation of the transposed convolutional layer.}
    \label{fig:basicConvTrans}
    \vspace{-2mm} 
\end{figure}

Then we convert the $\sum_{j=1}^{N}s_{ij}\phi_j[n]$ pattern to a neural network. We find the \textit{transposed convolutional layer} is a mathematically equivalent implementation. We first introduce the basic computation of a transposed convolutional layer in Figure~\ref{fig:basicConvTrans}. The $1$-D transposed convolutional layer has only $1$ input channel and $1$ output channel. The elements in input sequence $[+1,-1]$ are multiplied by a \textit{kernel}. The multiplication results are mapped to the output. The step between each multiplication result is determined by the \textit{stride} parameter.

\begin{figure}[t]
    \vspace{-2mm} 
    \centering
    \includegraphics[width=0.5\linewidth]{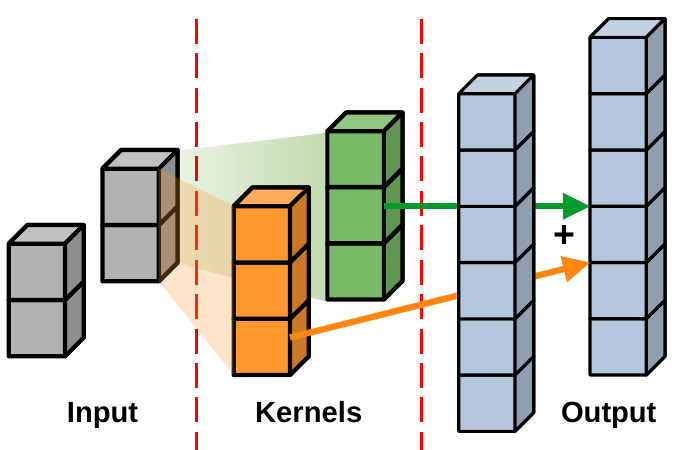}
    \caption{Diagram of the operation of multi-channel transposed convolutional layer.}
    \label{fig:multiChanConvTrans}
    \vspace{-5mm} 
\end{figure}

The transposed convolutional layer supports multiple input channels and multiple output channels~\cite{pytorch,tensorflow2015}. In Figure~\ref{fig:multiChanConvTrans}, we visualize the operation of the transposed convolutional layer with multiple input and output channels (both are $2$ in this figure). As illustrated here, each input channel will \textit{convolve} with a set of $2$ kernels, and the results are combined to generate one output channel. The calculation process of the transposed convolutional layer is the same as in one channel of Equation~\ref{eq:modulation_part}, \textit{if the kernel is set to the same as the real/imaginary parts of the basis functions, i.e., $\mathrm{Re}\{\phi_j[n]\}$ and $\mathrm{Im}\{\phi_j[n]\}$, and the stride is set to the samples per symbol, i.e., $L$ as in Equation~\ref{eq:modulation_sps}.}
\subsubsection{NN-defined Modulator Template} 
\begin{figure}[!htbp]
    \vspace{-2mm} 
    \centering
    \includegraphics[width=0.9\linewidth]{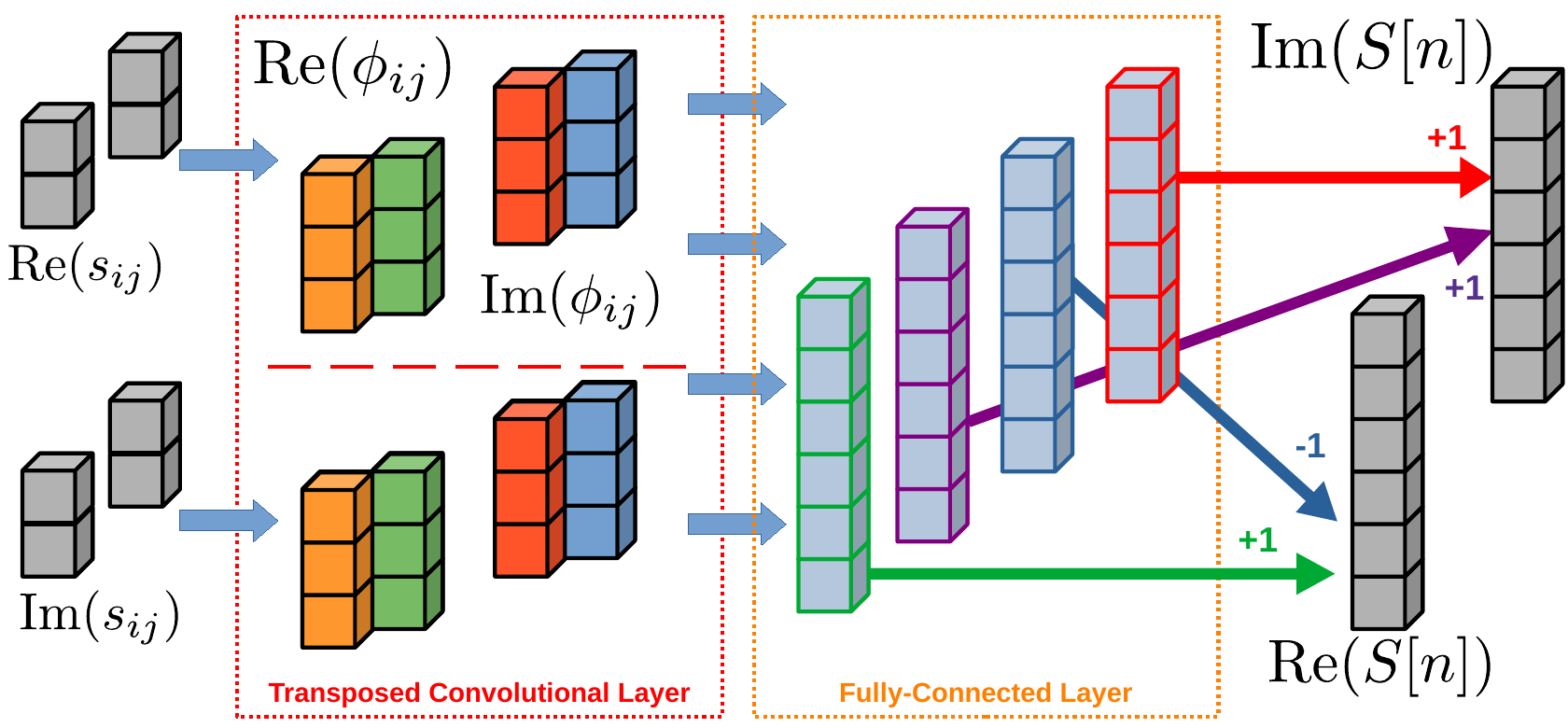}
    \vspace{-2mm} 
    \caption{Diagram of the template of NN-defined modulator. $0$-weight connections are omitted in the fully-connected layer.}
    \label{fig:TemplateNNModulator}
    \vspace{-3mm} 
\end{figure}

With the transposed convolutional layer, we can express the whole modulation process in Equation (\ref{eq:modulation_part}) as an NN-defined template modulator in Figure~\ref{fig:TemplateNNModulator}. The input channel comprises real and imaginary parts of the symbol vectors, which form two \textit{groups} of the transposed convolutional layer. The kernels of the transposed convolutional layer are determined by the basis functions. After that, a linear (fully-connected) layer is added to merge the four-channel outputs to generate the real and imaginary parts of the modulated signals. Its weight are set as $[+1, 0, 0, -1]$ and $[0, +1, +1, 0]$ according to the coefficients in Equation (\ref{eq:modulation_part}) as shown in Figure~\ref{fig:TemplateNNModulator}.

Thus, we begin with the mathematical foundation of the modulation process and derive a generalized modulation model. Subsequently, we show how to adapt the general model within our template for the NN-defined modulator. The universal template comprises a transposed convolutional layer followed by a fully-connected layer. By meticulously configuring the template for the NN-defined modulator, we can accomplish a range of modulation schemes.

\section{Instances of NN-defined Modulator Template} \label{sec:modulator}
With the NN-defined modulator template, we further study how to generate specific modulation schemes. 
\subsection{Common NN-defined Modulators}

\subsubsection{Single Carrier Amplitude/Phase Modulation}
For amplitude/phase modulation on the single carrier, the information is carried by the modulated signal's amplitude and/or phase. The most general case is quadrature amplitude modulation (QAM). For example, ZigBee \cite{ergen2004zigbee} adopts Offset-Phase-Shift Keying (O-QPSK) as its modulation scheme, which is a variant of QPSK or $4$-QAM scheme.

The QAM symbols are represented in a complex scalar as $s_k = \mathrm{Re}\{s_{k}\} + j\mathrm{Im}\{s_{k}\}$. The symbols pass a real-valued pulse-shaping filter to generate signals. Similar to equation~(\ref{eq:modulation_part}), we represent the I/Q signals as in
\begin{equation}
    \begin{aligned}
    S_I[n] &= \mathrm{Re}\{S[n]\}= \sum_k \mathrm{Re}\{s_{k}\}p[n-kL] \\
    S_Q[n] &= \mathrm{Im}\{S[n]\}= \sum_k \mathrm{Im}\{s_{k}\}p[n-kL]
    \end{aligned}
    \label{eq:QAM_IQ}
\end{equation}
where $p[n]$ represents the pulse-shaping filter, and $L$ is the number of samples per symbol. 

\begin{figure}[!htbp]
    \vspace{-2mm} 
    \centering
    \includegraphics[width=0.9\linewidth]{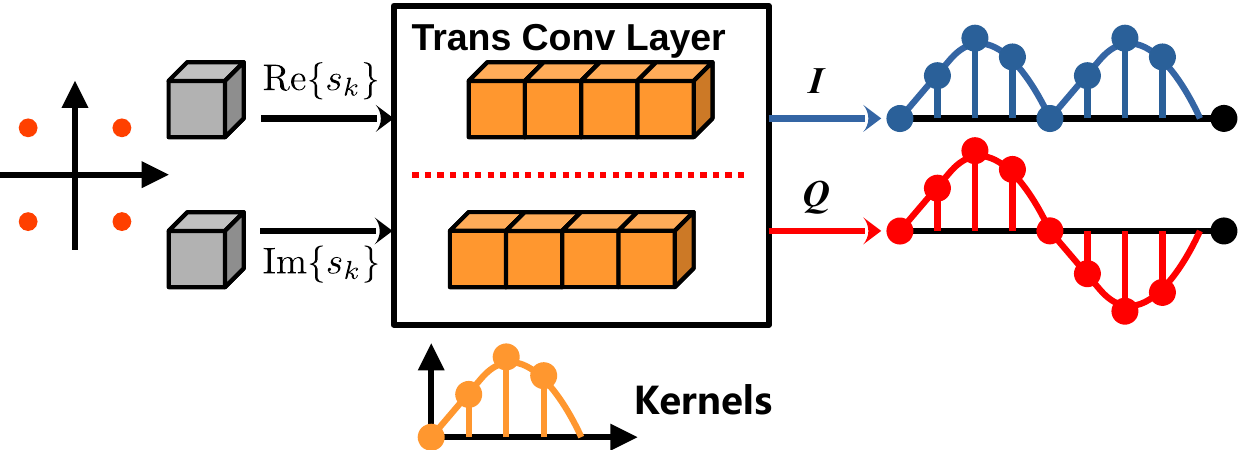}
    \caption{Diagram of a simplified NN-defined QPSK modulator with half-sine wave shaping filter.}
    \label{fig:QPSKMod}
    \vspace{-5mm} 
\end{figure}

Based on Equation~(\ref{eq:modulation_part}) and (\ref{eq:QAM_IQ}), when applying the NN-defined modulator template, we can configure the kernels of the transposed convolutional layer to be the values of shaping filter $p[n]$. This also implies the potential simplification of the template. If the shaping filter is real-valued, we can omit two channels of the transposed convolutional layer that correspond to the imaginary parts. We can also discard the fully-connected layer in the template because the output from the remaining $2$ output channels from the transposed convolutional layer directly forms the desired modulated signals. For better illustration, an NN-defined QPSK modulator with a half-sine wave shaping filter is depicted in Figure~\ref{fig:QPSKMod}. The output from the transposed convolutional layer is I/Q signals.

\subsubsection{Multicarrier Modulation}
We also extend our design for multicarrier modulation schemes, more specifically, the widely used OFDM scheme.

We consider a $N$-S.C. OFDM modulator as an example, of which the input symbol vector consists of $N$ elements, $s_{0}, s_{1}, \cdots, s_{N-1}$, that correspond to the components in the frequency domain. Thus, to get the signal samples $S[n]$, they are transformed to the time domain by performing an inverse Discrete Fourier Transform (IDFT) on the input $N$ elements, given as
\begin{equation}
    S[n]=\sum_{i=0}^{N-1}s_ie^{j2\pi ni/N}, \ 0\leq n \leq N-1.
    \label{eq:OFDM_IQ}
\end{equation}

The transformation can be interpreted as mapping the complex symbol vector $\mathbf{s} = [s_{0}, s_{1}, \cdots, s_{N-1}]$ of $N$ dimensions to signal $S[0],\cdots, S[N-1]$ with the basis functions set ${\phi_i[n]}$, which consists of $N$ functions in total, like $\phi_i[n]=e^{j(2\pi ni/N)}$. 

The OFDM scheme is consistent with the general case as in Equation~(\ref{eq:modulation_part}). As discussed in Section~\ref{sec:template}, the input to the NN-defined modulator consists of real and imaginary elements from the complex symbol vectors. They are divided into $2$ groups at the transposed convolutional layer. For each group, the kernels are set based on the real and imaginary parts of $e^{j2\pi ni/N}$. Then, the four-channel output from the transposed convolutional layer is fed into the fully-connected layer to generate the final In-phase and Quadrature signals.

\subsection{Protocol-specified NN-defined Modulators}
IoT protocol modulators may incorporate additional operations to enhance system reliability. For instance, ZigBee adopts an offset operation to the QPSK modulator by shifting the quadrature signals by half a symbol duration. OFDM systems used in WiFi adopt cyclic-prefix to improve robustness against multipath effects. Concurrently, some IoT protocols introduce intricate frame structures containing various fields for signaling. For example, WiFi frames typically encompass different signal fields. Although all these fields utilize the OFDM modulator, they may require different operations.

To address these additional operations, we draw inspiration from the inheritance feature in computer programming. The NN-defined modulators serve as the foundational component, and we attach operations to the temporal output from the base NN-defined modulator to generate the ultimate output signals. The attached processes are also achieved through operators supported by neural networks, allowing us to derive specialized NN-defined modulators for diverse protocols. We will discuss the protocol-specific NN-defined modulator in greater detail in Section~\ref{sec:eval}.

\begin{figure*}[t!]
    \vspace{-1mm}
     \centering
     \begin{subfigure}[b]{0.32\textwidth}
         \centering
         \includegraphics[width=0.9\linewidth]{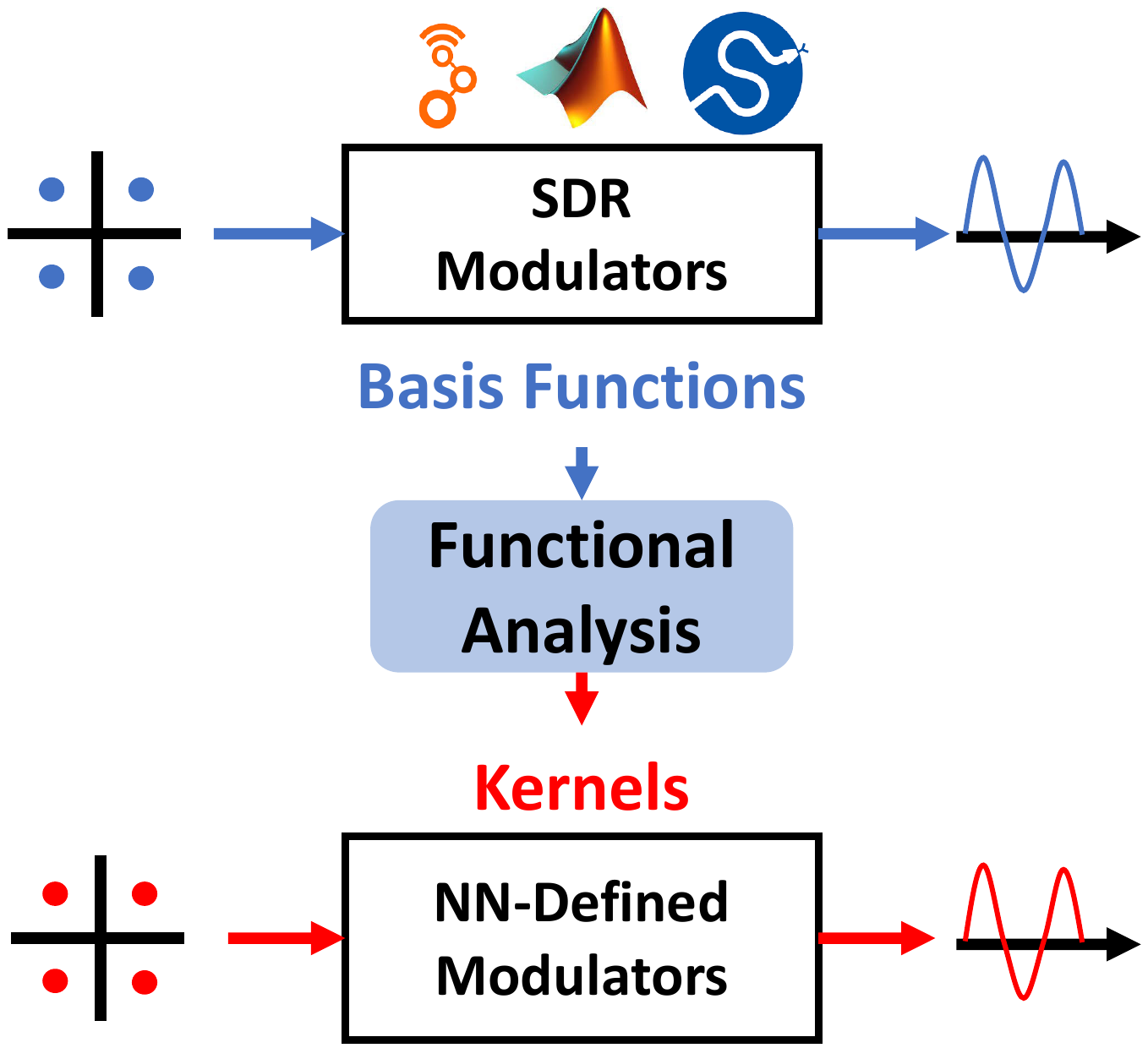}
         \caption{}
         \label{fig:ParaManual}
     \end{subfigure}
     \hfill
     \begin{subfigure}[b]{0.32\textwidth}
         \centering
         \includegraphics[width=0.9\linewidth]{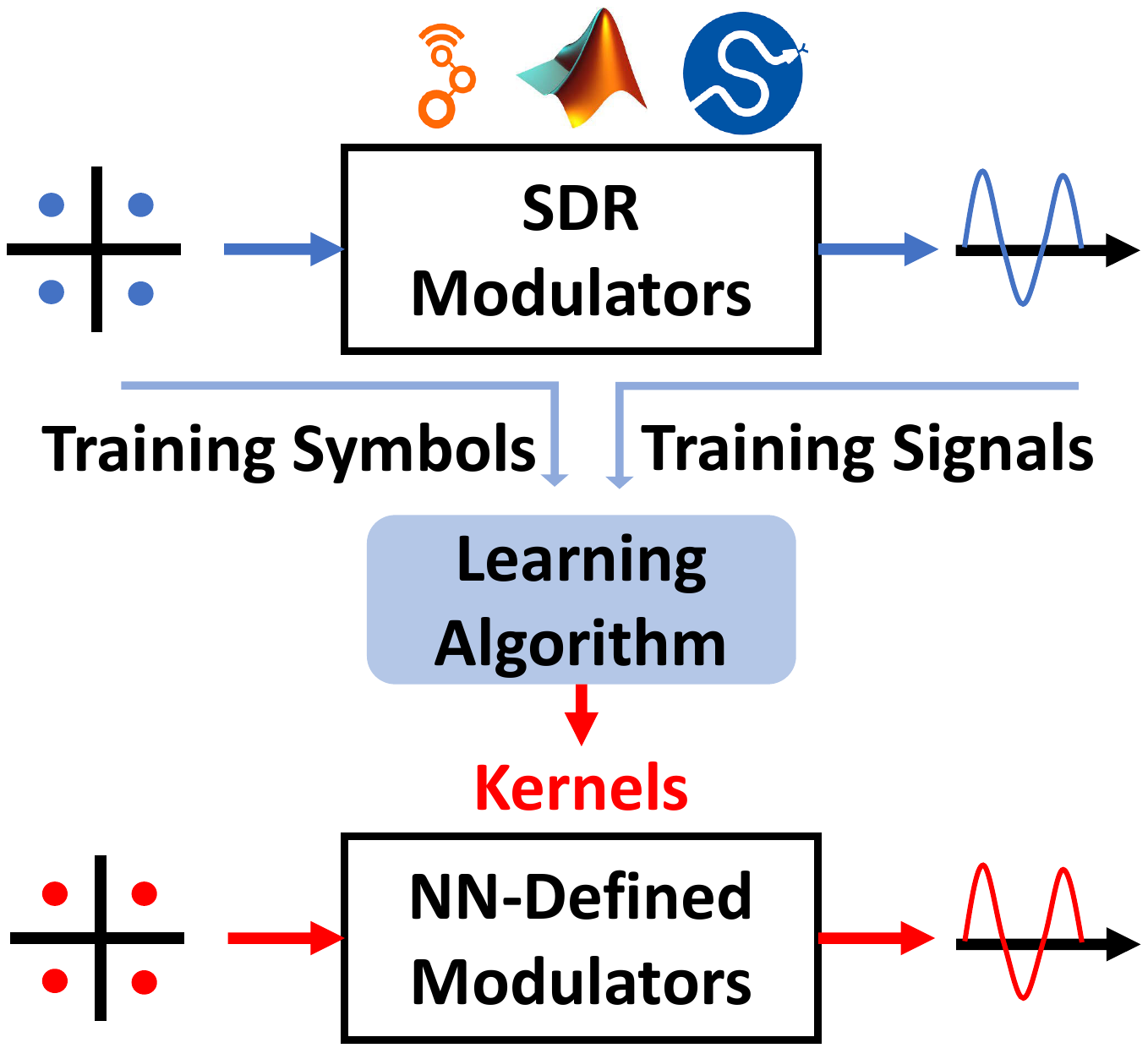}
         \caption{}
         \label{fig:ParaLearn}
     \end{subfigure}
     \hfill
     \begin{subfigure}[b]{0.32\textwidth}
         \centering
         \includegraphics[width=0.8\linewidth]{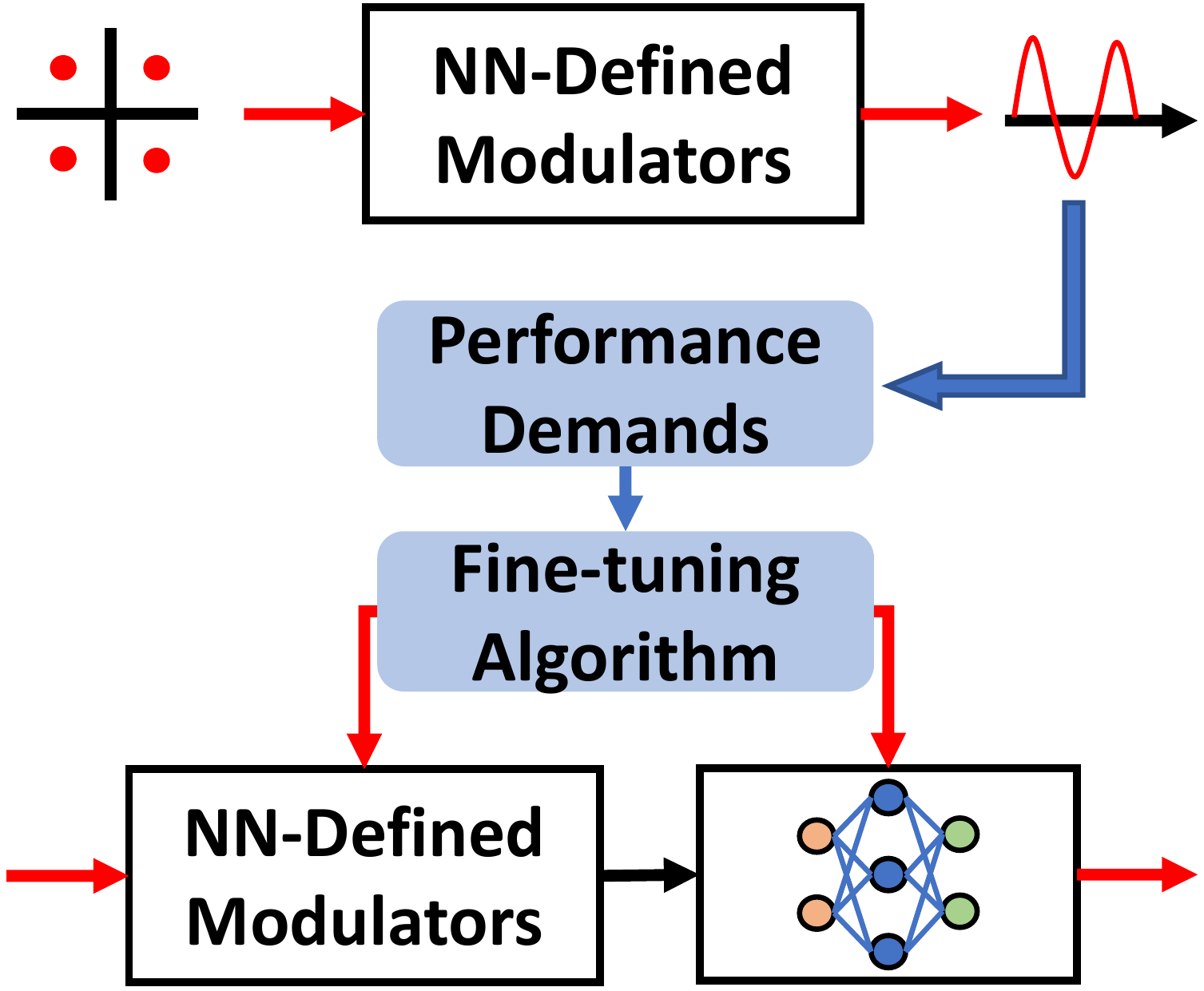}
         \caption{}
         \label{fig:ParaTune}
     \end{subfigure}
     \vspace{-2mm}
        \caption{Different approaches to configure kernels. (a) Manual setting with expert knowledge, (b) Learning from existing datasets, (c) Fine-tuning with other ML modules.}
        \label{fig:ParaConfig}
        \vspace{-5mm}
\end{figure*}

\section{Modulator Kernel Configuration}\label{sec:learning}
From the previous sections, we designed a template for the general modulation model, where the kernels of the template can be derived for a specific modulation scheme. In this section, we will discuss how to use the NN-defined modulator template to learn from signals whose analytical expression is unknown or fine-tune the NN-defined modulator to compensate for hardware distortion in practical systems. 

\subsection{Manual Setting with Expert Knowledge}
As shown in Figure~\ref{fig:ParaManual}, for a modulation scheme with a known analytical expression, communication experts can take a direct way to derive the kernels of the transposed convolutional layer as discussed in Section~\ref{sec:modulator}. It is an efficient and accurate way to construct signals in the NN-defined modulator, similar to the conventional Software modulator in the SDR development. 

\subsection{Learning from Dataset}
As shown in Figure~\ref{fig:ParaLearn}, for a signal with an unknown analytical expression or a non-expert developer, the kernels of the template can be derived by training the NN-defined modulator. For example, a non-expert developer who intends to shift an existing software radio to another platform can utilize the learning ability of the NN-defined modulator from the existing system to reconstruct the modulator, which will significantly ease the development complexity. One can treat it as a standard machine learning task to minimize the mean squared error. Thanks to the model-driven approach, the trained kernels imply a potential signal processing pipeline to mimic the target signal. 

More specifically, the training input has the dimension of $[Batch\_size, 2\times Symbol\_dimension,\ Sequence\_length]$, where $2\times Symbol\_dimension$ indicates the input is represented using the real and imaginary parts. And the training output has the size of $[Batch\_size, Signal\_length, 2]$, where the $2$ on the last dimension also indicates the real and imaginary parts of complex signals. There are $2\times Symbol\_dimension$ kernels to train in total. 

\begin{figure}[ht]
    \vspace{-3mm}
     \centering
    \includegraphics[width=0.6\linewidth]{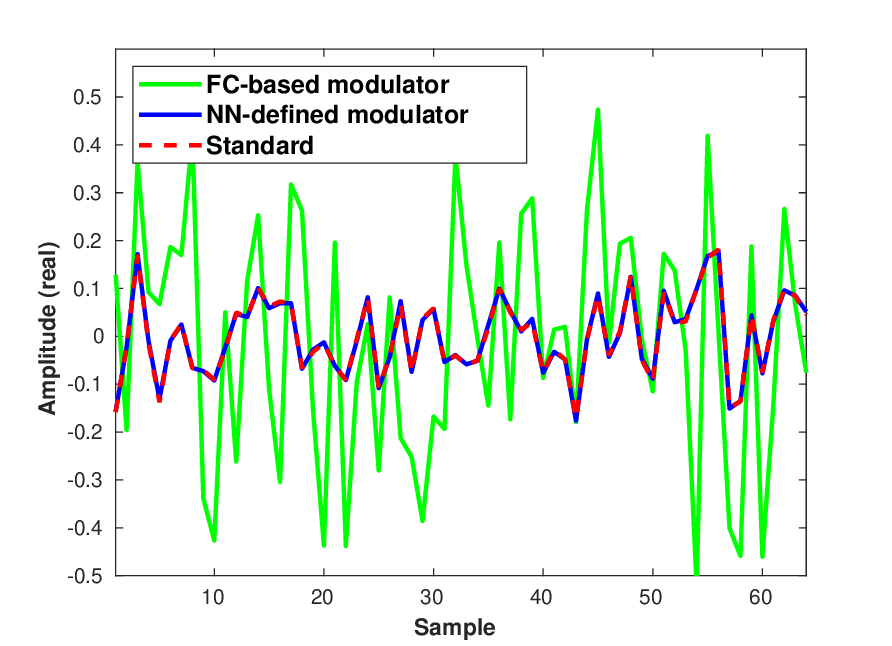}
     \vspace{-3mm}
     \caption{Waveform (In-phase) comparison of FC-based modulator, NN-defined modulator, and standard modulator for 64-S.C. OFDM scheme.}
     \label{fig:TemplateComparison}
     \vspace{-3mm}
\end{figure}

For demonstration, we apply the NN-defined modulator template to learn the $64$-S.C. OFDM scheme. The NN-defined OFDM modulator is trained with the same training settings as the example FC-based modulator in Section~\ref{sec:motivation} with the same dataset and training epochs. The training set contains $256$ different OFDM symbol sequences, each of which represents $128$ input complex symbols. The FC-based modulator is implemented with two fully-connected layers, with almost $\sim60000$ trainable parameters in total. We calculate the mean squared error between the modulated signals from two kinds of modulators and the standard signals on the training and test sets, respectively. Both the FC-based modulator and our NN-defined modulator have tiny errors on the training set. Our NN-defined modulator outperforms the FC-based modulator significantly on the test set. We plot signals generated from our NN-defined and the FC-based modulator in Figure~\ref{fig:TemplateComparison}. As in the figure, our NN-defined modulator can modulate the symbols correctly, while the FC-based modulator fails. The NN-defined modulator has much fewer parameters to train compared with the FC-based modulator, and the parameters are physically meaningful, which ensures that our NN-defined modulator is more reliable.

% As listed in Table~\ref{tab:TemplateTab}, 

% \begin{table}[ht]
%     \centering
%     \begin{tabular}{c c c}
%     \hline
%          Data set & MSE of FC-based & MSE of NN-defined \\
%     \hline
%     \hline
%         Training set & ~$1.5\times 10^{-6}$ & ~$2.3\times 10^{-5}$\\
%     \hline
%         Test set & ~$6.8 \times 10^{-2}$ &  ~$1\times 10^{-4}$\\
%     \hline
%     \end{tabular}
%     \caption{Mean Square Error (MSE) between standard signals and those modulated by FC-based and NN-defined modulator for 64-S.C. OFDM scheme.}
%     \label{tab:TemplateTab}
%     \vspace{-3mm}
% \end{table}

\subsection{Fine-tuning for Better Performance}
As shown in Figure~\ref{fig:ParaTune}, the NN-defined modulator can be combined with extra AI/ML models to fine-tune to meet specific performance demands. During the fine-tuning procedure, the kernels of the NN-defined modulators and parameters within the appended AI/ML module are adjusted to fulfill the goals. The fine-tuning process is an open design because the performance demands and the AI/ML extra modules are diverse. For better illustration, we discuss combining the proposed NN-defined modulator with additional AI/ML modules to handle the hardware distortion in the transmitter systems.

The modulated signals are processed at the RF front-end in the transmitter systems to send over the air. Due to the characteristics of the circuits, there is some non-linearity in the RF front-end hardware, which will introduce distortion to the output signal compared with the ideal output. One efficient approach to reduce the distortion effect is to apply the predistortion process to the modulated signals before feeding into the RF front-end~\cite{tarver2019dpd}. Here, we propose to use a neural network-based predistortion (NN-PD) module. Without loss of generality, we focus on the non-linearity introduced by the power amplifier.
 % And we demonstrate that the NN-defined modulator with NN-PD can be fine-tuned to handle the hardware distortion

\begin{figure}[ht]
    \vspace{-2mm}
    \centering
    \includegraphics[width=0.7\linewidth]{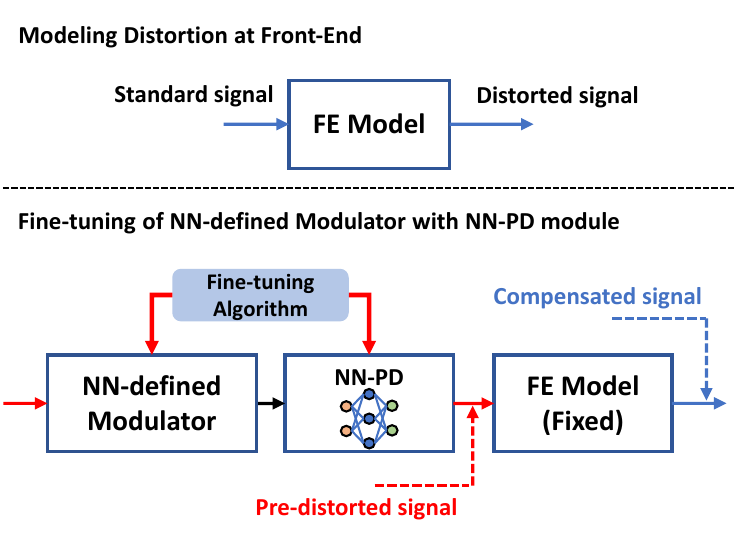}
    \vspace{-2mm}
    \caption{Diagram of Front-End model and NN-defined modulator with NN-PD module.}
    \label{fig:PAModel_NNPD}
    \vspace{-5mm}
\end{figure}

As illustrated in Figure~\ref{fig:PAModel_NNPD}, we first use a neural network, the front-end (FE) Model, to model the nonlinear behavior of the RF front-end. The FE model serves as the simulator of the RF front-end for the fine-tuning procedure. Next, we construct the NN-PD and insert it between the NN-defined modulator and the FE model. The predistorted signals from the NN-PD will pass the FE model, and the compensated signal is generated. The compensated signal is supposed to be as similar as possible to the ideal output signal. So, we set the training goal of our fine-tuning algorithm and tune the kernels in the NN-defined modulator and the parameters in NN-PD module while the parameters in the FE model are fixed. Once the fine-tuning procedure is finished, the NN-defined modulator and NN-PD module can generate predistorted signals, which can compensate for the non-linearity of the RF front-end.

\begin{figure}[!htbp]
    \vspace{-2mm}
    \centering
    \includegraphics[width=0.6\linewidth]{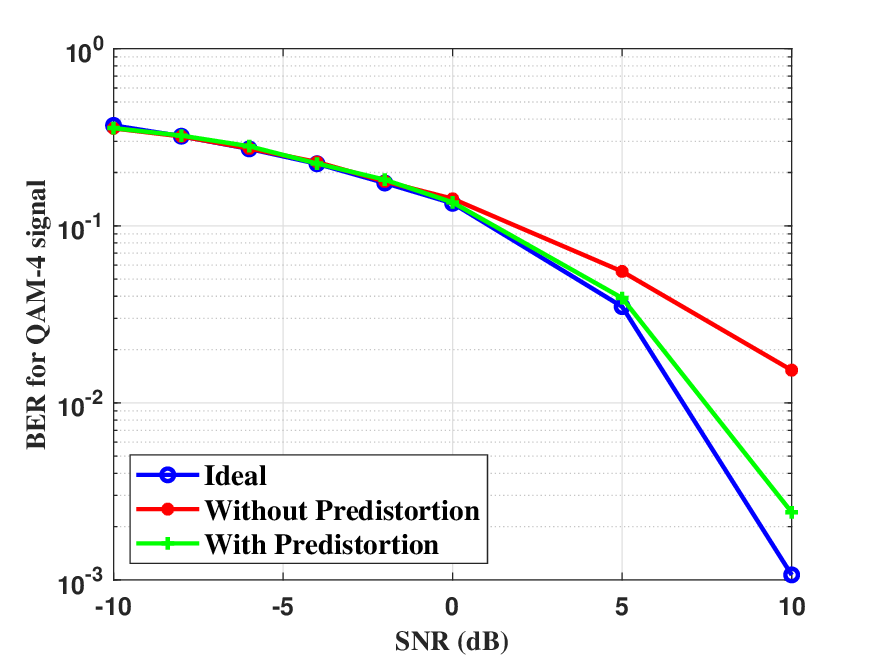}
    \vspace{-2mm}
    \caption{BER of NN-defined Modulator with NN-PD.}
    \label{fig:BER_NNPD}
    \vspace{-4mm}
\end{figure}

As for verification, we compare the Bit Error Rate (BER) performance of QAM-modulated signals with predistortion and those without predistortion. The simulation is conducted in additive white Gaussian noise (AWGN) channel. We plotted the BER curves in Figure~\ref{fig:BER_NNPD}. The BER curve of the ideal signals is also visualized as the baseline. Furthermore, we conduct Error Vector Magnitude~(EVM) test on the modulated signals. EVM can be evaluated by a percentage scale that reflects the deviation of the modulated and standard constellations. We measure the root mean squared EVM of the signals at different SNR levels. The results are illustrated in Table~\ref{tab:dpd}. When the SNR is low (SNR$<0$dB), the noise is dominant in such conditions, so all three signals suffer from noisy environments, resulting in high error rates and high EVM. However, when SNR is relatively high (SNR$>0$dB), the distortion effect of the RF front-end is more significant than the noise. Hence, the signals with predistortion perform much better than those without predistortion because the hardware distortion is reduced. However, the compensation is imperfect, so the error rates and EVMs of the predistorted signals are still slightly larger than the ideal signals. The above results indicate the great potential that the proposed NN-defined modulator can be integrated with other AI/ML modules and deliver better performance. 

\begin{table}[ht]
\newcommand{\tabincell}[2]{\begin{tabular}{@{}#1@{}}#2\end{tabular}}
    \centering
    \begin{tabular}{c c c c}
        \hline
        \tabincell{c}{} & \tabincell{c}{SNR=-10dB} & \tabincell{c}{0dB} & \tabincell{c}{10dB}\\
        \hline
        \hline
        EVM of ideal signals  & 65.9\% & 31.2\% & 15.4\% \\
        \hline
        w/ predistortion & 66.6\% & 32.1\% & 15.7\%  \\
        \hline
        w/o predistortion & 79.5\% & 33.4\% & 21.7\%  \\
        \hline
    \end{tabular}
    \caption{Root mean squared EVM of ideal modulated signals, signals with predistortion, and signals without predistortion.}
    \label{tab:dpd}
    \vspace{-5mm}
\end{table}

\section{Modulator with Portability}\label{sec:portability}
% In this section, we discuss the key enabler for the proposed NN-defined modulator, highlighting the practical portability compared with other designs.

We first highlight better portability of NN-defined modulators compared with conventional software radio systems by demonstrating the pipeline of the software modulator to generate signal samples. The software radio relies on some libraries which contain signal processing operations. We choose SciPy~\cite{scipy}, a scientific computing library in Python, and GNURadio~\cite{GNURadio}, the recognized software radio library, as case studies. Moreover, without loss of generality, we consider the QAM with Root Raised Cosine~(RRC) pulse shaping filter as an example. It requires two major steps for modulation: upsampling and pulse-shaping filtering. We list the corresponding implementation for the GNURadio-based and the SciPy-based modulators in Table~\ref{tab:SDRimplQAM}. As shown in the table, although the two kinds of implementations share the same pipeline, the functions used are quite different, which requires the developer to master new development tools for smooth conversion. Besides, we also notice that GNURadio provides several predefined shaping filter blocks, like \texttt{Root Raised Cosine Filter(rrc\_fir)} for quick usage. In contrast, SciPy does not provide such predefined functions, so we need to configure the filter manually, which also increases the difficulty of porting.

\begin{table}[ht]
\newcommand{\tabincell}[2]{\begin{tabular}{@{}#1@{}}#2\end{tabular}}
    \centering
    \begin{tabular}{c c c }
        \hline
        \tabincell{c}{Operations} & \tabincell{c}{GNURadio} & \tabincell{c}{SciPy} \\
        \hline
        \hline
        Upsampling  & \texttt{interp\_fir} & \texttt{scipy.interpolate}  \\
        \hline
        Filtering   & \texttt{rrc\_fir} & \texttt{scipy.convovle}    \\
        \hline
    \end{tabular}
    \caption{Operations for QAM modulator in different toolkits.}
    \label{tab:SDRimplQAM}
    \vspace{-5mm}
\end{table}

\subsection{Framework-Independent NN-defined Modulators}
\textbf{Framework-dependent design} implies that the NN-based modulators depend on unique functions or models provided by specific machine learning frameworks. Although machine learning frameworks offer some mathematical functions that can be employed to develop customized neural network layers for modulation tasks, as seen in NVIDIA Sionna~\cite{hoydis2022sionna} based on TensorFlow, the customized neural network modulator remains reliant on the development framework. To exemplify, we demonstrate the implementation details of the Sionna-based QAM modulator. Sionna employs the built-in operations and encapsulates them into the customized neural network layers, \texttt{Upsampling} and \texttt{Filter}, to emulate the functions as in the conventional pipeline. \texttt{Upsampling} layer applies \texttt{tf.pad} and some dimensional operations like \texttt{tf.expand\_dims} to insert zeros between symbols, and \texttt{Filter} layer applies \texttt{tf.math.convolve} which takes the upsampled symbol sequences and filter taps as input to generate the modulated signals. We compare the availability of mathematical functions used in the Sionna modulator across other mainstream ML frameworks. The results are presented in Table~\ref{tab:NNSionnaimpl}. Although there are similar functions such as \texttt{pad} and \texttt{convolve}, the direct transition among different frameworks is still hard. Consequently, the framework-dependent modulator can be ported to platforms running the same framework, but it is challenging to deploy it on platforms operating with different frameworks.

\begin{table}[ht]
\vspace{-2mm}
\newcommand{\tabincell}[2]{\begin{tabular}{@{}#1@{}}#2\end{tabular}}
    \centering
    \begin{tabular}{c c c c}
        \hline
        \tabincell{c}{} & \tabincell{c}{Tensorflow} & \tabincell{c}{PyTorch} \\
        \hline
        \hline
        \multirow{2}{*}{\tabincell{c}{NN-defined}}
        & \texttt{Conv1DTranspose} & \texttt{ConvTranspose1d} \\
        \cline{2-3}
        & \texttt{Linear} & \texttt{Linear}    \\
        \hline
        \multirow{3}{*}{\tabincell{c}{Sionna}}
        & \texttt{pad}  & \texttt{pad}+\texttt{concatenate} \\
        \cline{2-3}
        & \texttt{expand\_dims} & \texttt{unsqueeze} \\
        \cline{2-3}
        & \texttt{convolve} & \texttt{convolve}    \\
        \hline
    \end{tabular}
    \caption{Original operations and converted ONNX operators in our NN-defined and Sionna QAM modulator.}
    \label{tab:NNSionnaimpl}
    \vspace{-4mm}
\end{table}

\textbf{Framework-independent design} means our NN-based modulators are implemented by the share functions or models by various machine learning frameworks. Unlike NVIDIA Sionna, which constructs customized layers fro modulators, our NN-defined modulators utilize the fundamental neural network layers that are considered basic components of existing machine learning frameworks. More specifically, the transposed convolutional layer and the fully connected layer are generally supported by various frameworks~\cite{tensorflow2015,pytorch,chen2015mxnet}. Although the layer names vary~(Table~\ref{tab:NNSionnaimpl}), they share the same functionalities, which ensures that the proposed NN-defined modulator can be a framework-independent design. 

\begin{figure}[t]
     \centering
     \begin{subfigure}[b]{0.35\linewidth}
         \centering
         \includegraphics[width=0.7\linewidth]{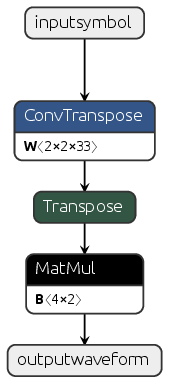}
         \caption{}
         \label{fig:PortabilityCase}
     \end{subfigure}
     \begin{subfigure}[b]{0.55\linewidth}
         \centering
         \includegraphics[width=0.9\linewidth]{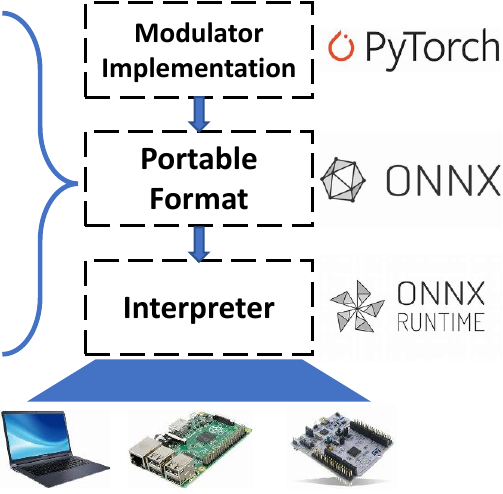}
         \caption{}
         \label{fig:PortabilityDiagram}
     \end{subfigure}     
     \caption{(a) Example converted ONNX format of a QAM NN-defined modulator, (b) Diagram of development and deployment of NN-defined modulators.}
     \vspace{-5mm}
\end{figure}

We utilize the ONNX~\cite{ONNX} as an intermediate framework to ensure the interoperability. ONNX is an open ecosystem for technology companies and research organizations to store and import neural network models onto different frameworks. ONNX defines a common set of operators that contains the fundamental layers of neural network models, including the transposed convolutional layer and the fully-connected layer used in our design. As a validation, we visualize the graph of the ONNX model of our NN-defined modulator template in Figure~\ref{fig:PortabilityCase}. As depicted in the figure, the transposed convolutional layer operator is \textit{ConvTranspose}, and the fully-connected layer is represented by the \textit{MatMul} operator. Almost all mainstream machine learning frameworks support conversions between their native models and ONNX ones. It is also worth noting that porting customized neural network layers to ONNX models demands significant effort. Consequently, the custom layers in NVIDIA Sionna are challenging to convert to ONNX models, while our NN-defined modulator built upon fundamental layers exhibits better interoperability across different frameworks.

\subsection{Seamless Acceleration}
As previously illustrated, IoT gateway hardware platforms provide acceleration capabilities to expedite the execution of neural network models. The proposed NN-defined modulator is constructed based on fundamental neural network layers that are generally supported and well-optimized for execution on various hardware platforms. Therefore, the NN-based modulator can leverage these capabilities to enhance efficiency, speeding up the modulation process. 

A typical development and deployment workflow for the proposed NN-defined modulators is depicted in Figure~\ref{fig:PortabilityDiagram}. The prototype of the NN-defined modulators can be developed in mainstream machine learning frameworks, such as PyTorch. Then, NN-defined modulators are converted to a portable ONNX format for improved interoperability across different platforms. Deploying ONNX models requires a compiler, such as ONNX runtime~\cite{onnxruntime} or Apache TVM\cite{TVM}. Using ONNX runtime as an example, it can utilize different accelerator backends. Numerous accelerator backends have been developed by the community. For instance, it can employ Nvidia GPU~\cite{Cuda} on GPU-equipped systems, Arm ACL~\cite{ACL} for Arm SoC platforms, and OpenVINO~\cite{OpenVINO} for Intel x86 platforms. Therefore, the NN-defined modulator can be seamlessly accelerated on various platforms.

\section{Evaluation}\label{sec:eval}
\subsection{Implementation}
\subsubsection{Framework and hardware}
We design the NN-defined modulator in PyTorch \cite{pytorch} with \texttt{ConvTranspose1d} and \texttt{Linear} layers. Once the NN-defined modulators are ready for port, we convert the modulators into ONNX format. We port the ONNX NN-defined modulator to Nvidia Jetson Nano~\cite{jetson} and Raspberry Pi~\cite{Rpi} for verification. Both devices support the ONNX runtime, and Jetson Nano is equipped with a GPU, which can be used to accelerate the execution of the ONNX NN-defined modulators.

\begin{figure}[t]
    \centerline{\includegraphics[width=0.6\linewidth, keepaspectratio=true]{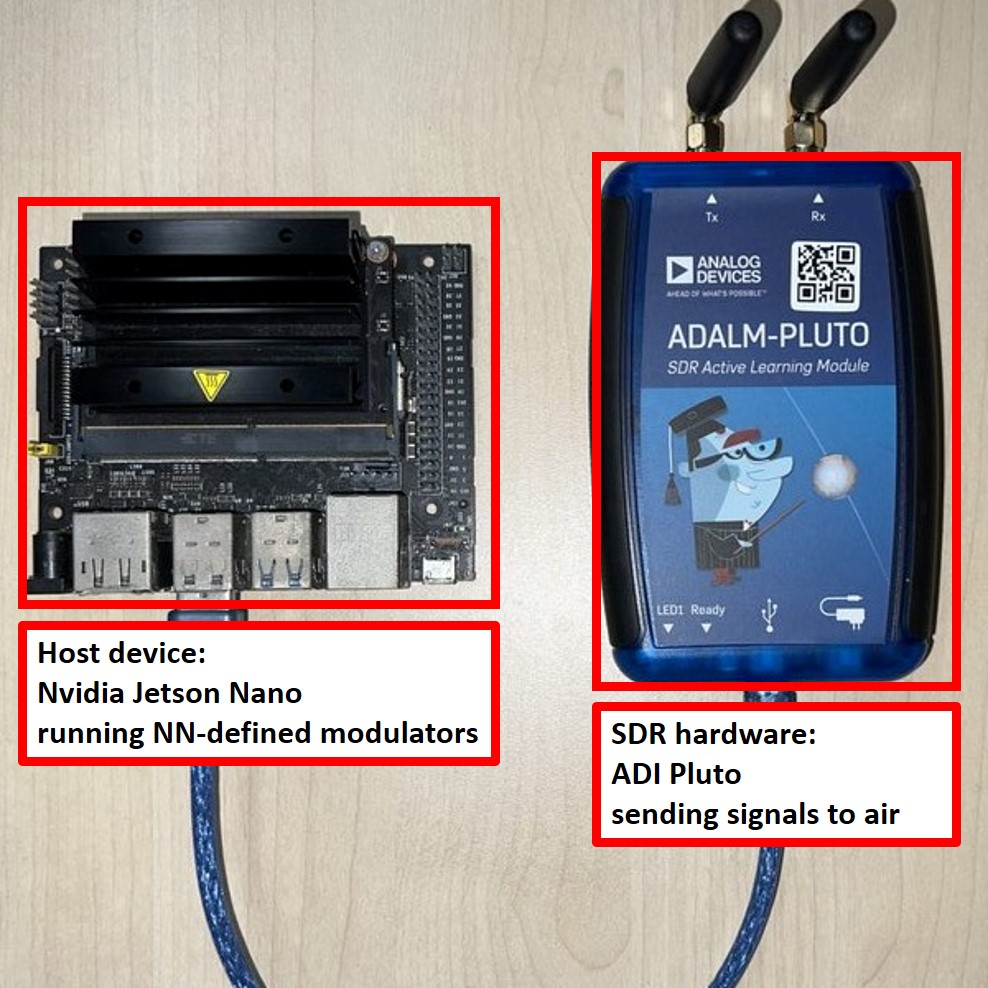}}
    \caption{Prototype of NN-defined modulator. Left box: Nvidia Jetson Nano as the host device for the NN-defined modulator. Right box: ADI Pluto SDR as SDR hardware.}
    \label{fig:TestbedOverview}
    \vspace{-5mm} 
\end{figure}

Besides, we also implement an NN-defined modulator prototype. We connect the host (Nvidia Jetson Nano) running the NN-defined modulator with the SDR hardware (ADI Pluto SDR~\cite{ADI}) as shown in Figure~\ref{fig:TestbedOverview}. We use this prototype to transmit the modulated signals over the air.

\subsubsection{Modulation schemes}
Without loss of generality, we choose several typical schemes for 1) PAM-2 with the rectangular filter, 2) QPSK with the half-sine wave filter, 3) $16$-QAM with RRC filter for amplitude/phase modulation, and 4) $64$-S.C. OFDM scheme for multicarrier modulation. We use MATLAB Signal Processing Toolbox~\cite{MatlabSP} to generate the symbols and the signals as for training sets. When evaluating the efficiency and portability, we select $16$-QAM modulator with RRC filter as the example. The conventional SDR modulators for $16$-QAM with RRC filter are implemented with signal processing libraries, GNURadio \cite{GNURadio} on x86 laptop, and SciPy \cite{scipy} on Nvidia Jetson Nano, as the baselines. For comparison, we also implement a $16$-QAM modulator with RRC filter using Nvidia Sionna \cite{hoydis2022sionna} on the x86 laptop for comparison.

\subsection{Signal Quality of NN-defined Modulator}
\subsubsection{Trained kernels in NN-defined modulators}
As discussed in Section~\ref{sec:learning}, the kernels within the NN-defined modulator can be trained with training sets. Here, we use $16$-QAM with RRC filter and $64$-S.C. OFDM scheme as examples to analyse the trained kernels. 

\begin{figure}[!htbp]
    \vspace{-3mm}
     \centering
     \begin{subfigure}[b]{0.45\linewidth}
         \centering
         \includegraphics[width=1.05\linewidth]{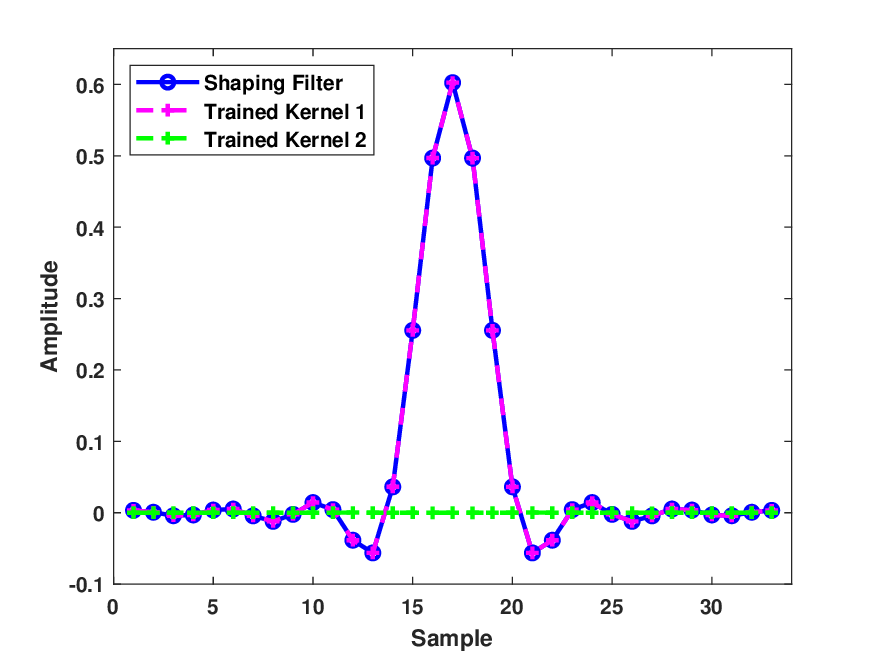}
         \caption{}
         \label{fig:Weight_QAM}
     \end{subfigure}
     \begin{subfigure}[b]{0.45\linewidth}
         \centering
         \includegraphics[width=1.1\linewidth]{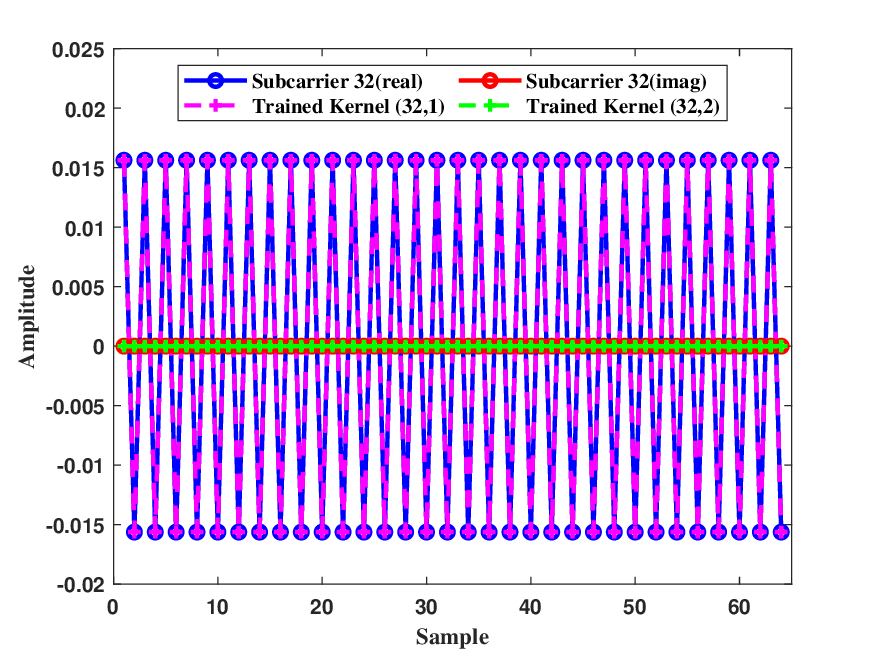}
         \caption{}
         \label{fig:Weight_OFDM}
     \end{subfigure}  
     \caption{Trained kernels from NN-defined modulators for (a) QAM with RRC filter, (b) $64$-S.C. OFDM.}
    \vspace{-5mm}
\end{figure}

For the $16$-QAM scheme with RRC shaping filter, the input symbol is $1$-dimensional, so there are $2$ kernels trained. According to the analysis in Section~\ref{sec:modulator}, the trained kernels are supposed to be the real and imaginary parts of the shaping filter. We visualize trained kernels and the original RRC shaping filter in Figure~\ref{fig:Weight_QAM}. One of the trained kernels is nearly identical to the original shaping filter. The other one is almost zero-valued, which is consistent with the zero-valued imaginary parts of the shaping filter. 

For the $64$-S.C. OFDM scheme, there are $2\times 64$ kernels trained. According to the analysis in Section~\ref{sec:modulator}, the kernels are supposed to be the real and imaginary parts of the subcarrier functions, i.e., $e^{j\frac{2\pi i n}{64}}$. We also visualized a pair of the trained kernels in Figure~\ref{fig:Weight_OFDM}. And these two kernels are the same as real and imaginary parts of the standard subcarrier $e^{j\frac{2\pi \times 32 n}{64}}$. The NN-defined OFDM modulators share the same conclusion that the trained kernels perfectly match the signal processing pipeline in conventional modulators.

\subsubsection{Transmission performance in AWGN channel}
\begin{figure}[!htbp]
    \vspace{-1mm} 
    \centerline{\includegraphics[width=0.7\linewidth, keepaspectratio=true]{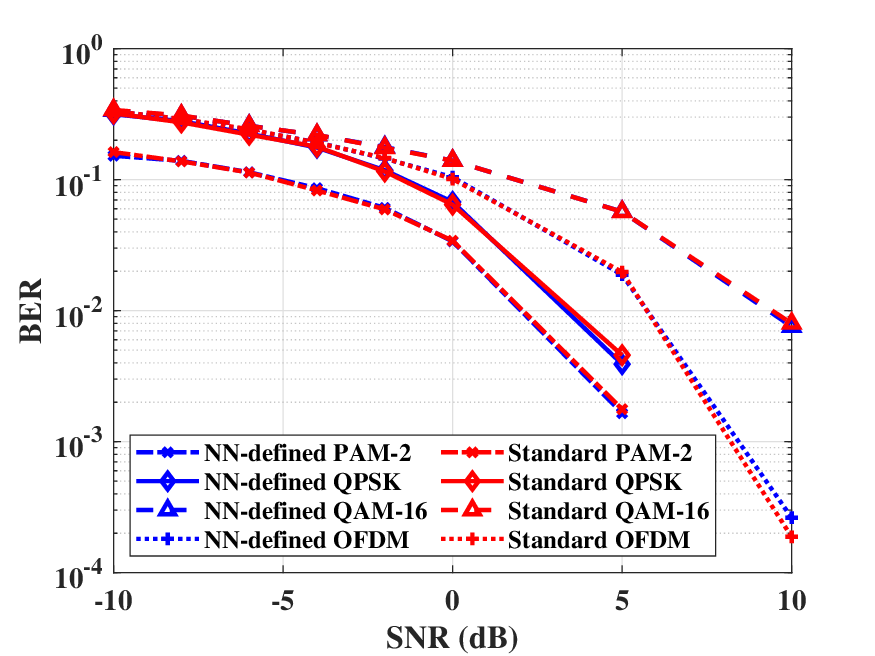}}
    \caption{The BER performance of NN-defined modulator compared with standard modulators.}
    \label{fig:BER}
    \vspace{-5mm} 
\end{figure}

We apply the trained NN-defined modulators to generate signals and pass the signals in the additive white Gaussian noise (AWGN) channel to verify the transmission performance. And we plot the Bit Error Rate (BER) curves in Figure~\ref{fig:BER}. Meanwhile, The BER curves of the signals from standard modulators in MATLAB are also plotted as the baseline. As illustrated in the figure, the NN-defined modulators for the selected modulation schemes can modulate the symbols correctly so that the modulated signals can achieve the same error performance as standard modulators in AWGN channels.

\subsection{Efficiency and Portability}
\subsubsection{Efficiency improvement}
\begin{figure}[!htbp]
    \vspace{-2mm}
    \centerline{\includegraphics[width=0.7\linewidth, keepaspectratio=true]{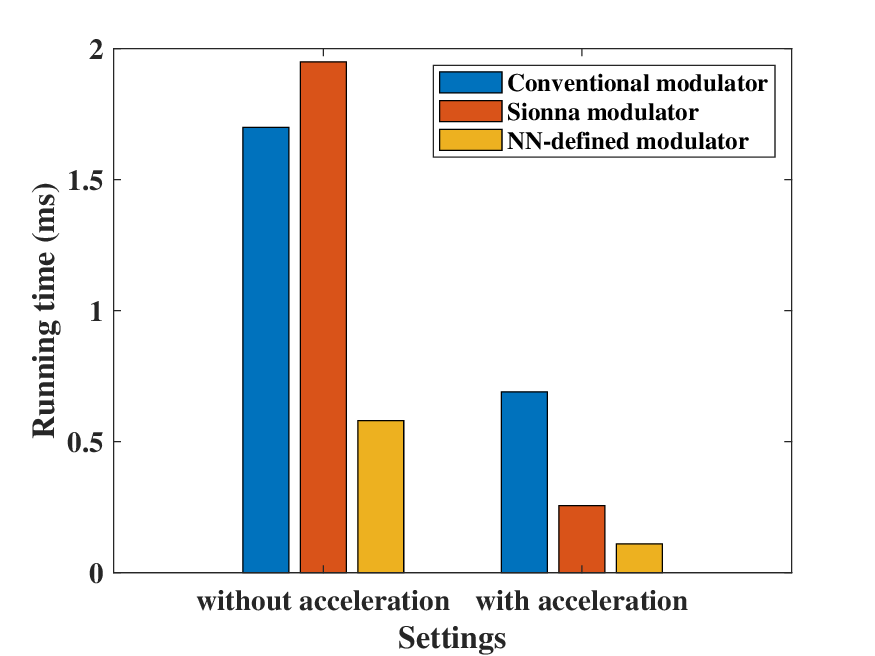}}
    \vspace{-3mm}
    \caption{Running time of different implementations.}
    \label{fig:Acceleration}
    \vspace{-3mm} 
\end{figure}

To verify the efficiency improvement of the NN-defined modulator, we measure the running time of the conventional SDR QAM modulator, the NN-defined QAM modulator, and the Nvidia Sionna QAM modulator on an x86 laptop and compare these time recordings in Figure~\ref{fig:Acceleration}. All the QAM modulators modulate a batch consisting of $32$ symbol sequences with $256$ symbols.

When all three modulators run without acceleration, it takes $0.58$ms for our design to finish, which is much faster than the conventional ($1.7$ms) and Sionna modulator ($1.9$ms). Our NN-defined modulator applies the fundamental neural network layers that are well-optimized so that it performs better than the customized neural network layers implemented in Nvidia Sionna as well as the conventional SDR modulator.
% Conventional QAM modulators need two steps, upsampling and pulse-shaping filtering \cite{goldsmith2005wireless}, and Sionna re-implements these operations with customized neural layers.

The NN-defined and Sionna modulators support hardware acceleration thanks to the neural network implementation. We measure their running time with acceleration enabled. We also implement an accelerated QAM modulator with cuSignal~\cite{CuSignal}. The NN-defined modulator and Sionna modulator execute much faster than without acceleration. The running time of the NN-defined modulator is reduced to $0.059$ms from $0.58$ms, which is $28$ times faster than the conventional SDR modulators without acceleration, even $10$ times faster than the implementation using cuSignal. For the Sionna modulator, the running time is also reduced to $0.25$ms. Both our NN-defined modulator and Sionna modulator run faster than the conventional modulator. These results prove that the NN-defined modulator can significantly improve efficiency compared with conventional SDR modulators. 

\subsubsection{Portability} 

\textbf{Porting among platforms.} 
As aforementioned in Section~\ref{sec:portability}, porting the conventional SDR implementations and Sionna-based one from one platform to another requires considerable effort. Here, we focus on the portability of our NN-defined modulator. Following the development diagram of the NN-defined modulators, we first implement the NN-defined QAM modulator in PyTorch and convert it into the ONNX model. We list the converted operations in the ONNX framework in Table~\ref{tab:QAM_ONNX}. The transposed convolutional layer~(\texttt{torch.ConvTranspose1d}) and the linear layer~(\texttt{torch.Linear}) are widely supported so that they can be converted to the portable format. 

\begin{table}[ht]
\newcommand{\tabincell}[2]{\begin{tabular}{@{}#1@{}}#2\end{tabular}}
    \centering
    \begin{tabular}{c c c}
        \hline
        \tabincell{c}{Implementations} & \tabincell{c}{PyTorch layer} & \tabincell{c}{ONNX operator} \\
        \hline
        \hline
        \multirow{2}{*}{\tabincell{c}{NN-defined}}
        & \texttt{ConvTranspose1d} & \texttt{ConvTranspose} \\
        \cline{2-3}
        & \texttt{Linear} & \texttt{MatMul}    \\
        \hline
    \end{tabular}
    \caption{Original operations and converted ONNX operators in the NN-defined modulator.}
    \label{tab:QAM_ONNX}
    \vspace{-5mm}
\end{table}

\begin{figure}[!htbp]
    \vspace{-2mm}
    \centering
    \begin{subfigure}[b]{0.23\textwidth}
        \includegraphics[width=1.05\textwidth, keepaspectratio=true]{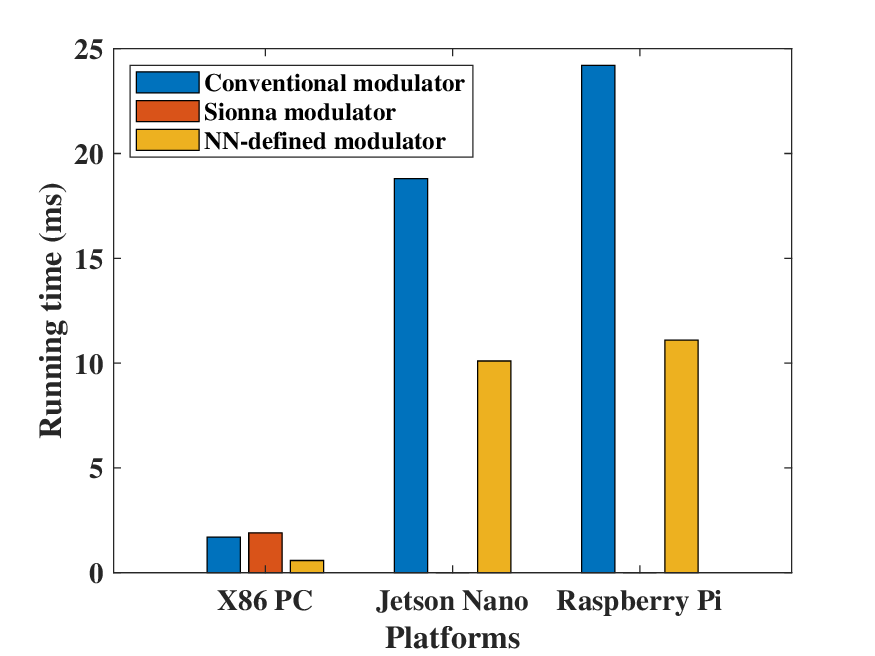}
        % \caption{Porting a QAM-4 modulator deveolped on PC to an ARM-based system}
        \caption{}
        \label{fig:ONNX_portability}
    \end{subfigure}
    %\hfill
    \begin{subfigure}[b]{0.23\textwidth}
        \includegraphics[width=1.05\textwidth, keepaspectratio=true]{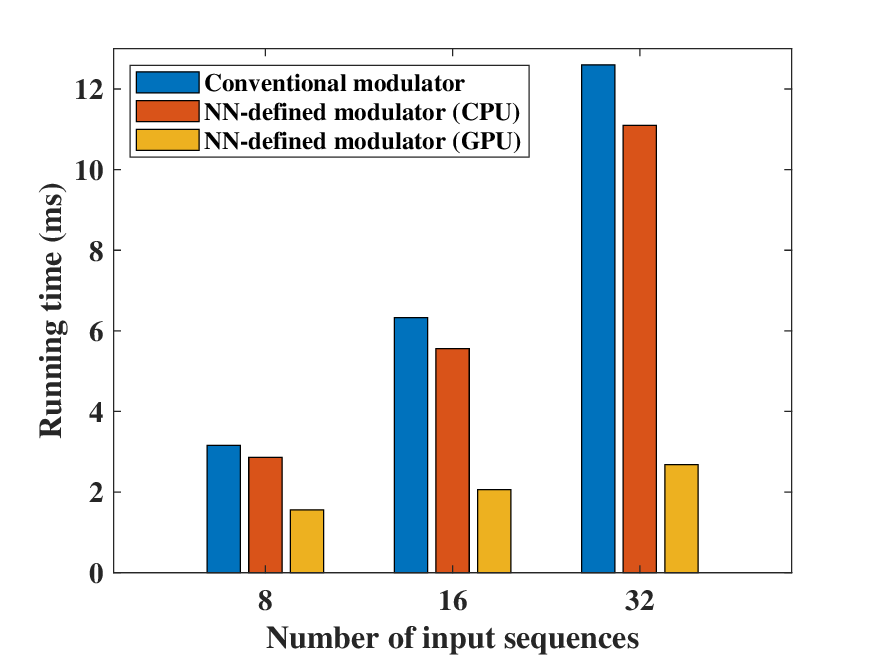}
        % \caption{Porting a QAM-4 modulator deveolped on PC to an ARM-based system}
        \caption{}
        \label{fig:ONNX_backends}
    \end{subfigure}
    \vspace{-2mm}
    \caption{(a) Running time on different platforms of x86 PC, Nvidia Jetson Nano, Raspberry Pi. (b) Acceleration evaluation on the target platform of Nvidia Jetson Nano.}
    \label{fig:ONNX_evaluation}
    \vspace{-4mm}
\end{figure}

\textbf{Performance on different platforms.} We now deploy the ONNX NN-defined QAM modulator on embedded computers such as Nvidia Jetson Nano and Raspberry Pi. Figure~\ref{fig:ONNX_portability} illustrates the running time on different platforms. Sionna modulator fails to be ported because the customized layers are hard to be transformed into ONNX models. Although the running time of the NN-defined modulator on embedded systems is longer than that on the x86 laptop, we successfully port our NN-defined modulators to different platforms.

The following evaluation demonstrates the acceleration capability of the target platform. We configure the ONNX NN-defined QAM modulator on Nvidia Jetson Nano to run with GPU acceleration as discussed in Section~\ref{sec:portability}. We compare the running time of the conventional modulator and our NN-defined modulator modulate symbol batches of different sizes. The evaluation results are visualized in Figure~\ref{fig:ONNX_evaluation}. We can observe a considerable efficiency improvement compared with the conventional modulator as well as the CPU-only NN-defined modulator. Moreover, the efficiency of our NN-defined modulator is still much better than the conventional modulator implemented with an accelerated signal processing library. More specifically, when the number of input symbol sequences is $32$, the accelerated NN-defined modulator is $4.7$ times faster than the conventional modulator and even $2.5$ times faster than the accelerated modulator. These results showcase that we can easily run the NN-defined modulators on target platforms with acceleration capability.

\subsection{Application in IoT Technologies}
The proposed NN-defined modulators are employed to generate protocol-compliant signals, showcasing representative use cases in IoT gateways.

\subsubsection{ZigBee-compliant Signals}

\begin{figure}[!htbp]
    \vspace{-2mm} 
    \centerline{\includegraphics[width=0.8\linewidth, keepaspectratio=true]{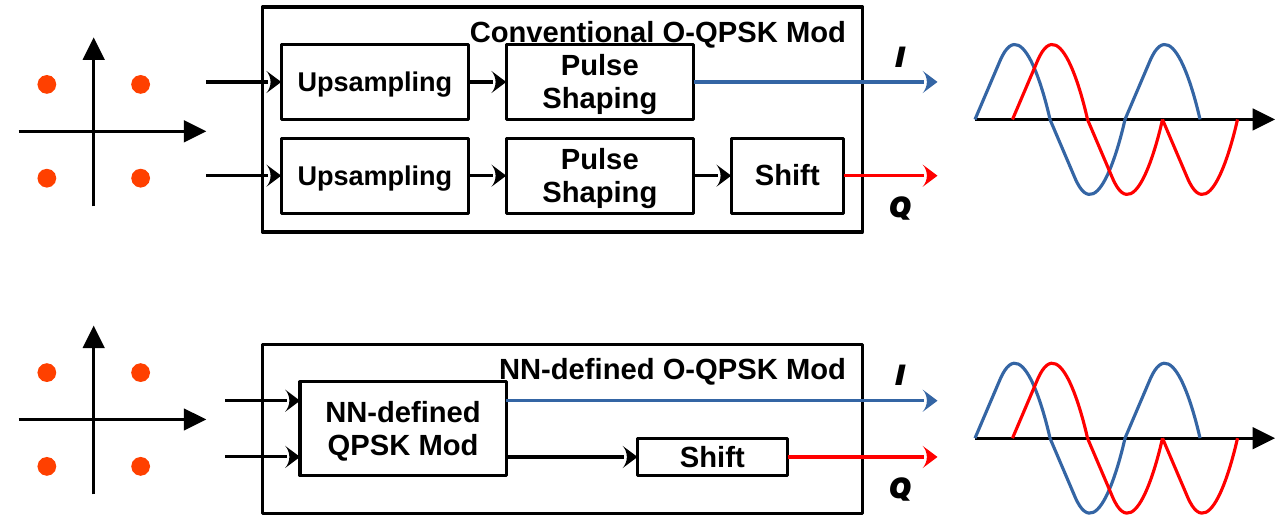}}
    \caption{Diagram of conventional O-QPSK modulator (top) and the NN-defined O-QPSK modulator (bottom).}
    \label{fig:ZigBeeMod}
    \vspace{-4mm} 
\end{figure}

ZigBee \cite{ergen2004zigbee}, developed based on IEEE 802.15.4 \cite{ieee802154}, utilizes Offset-QPSK, a variant of amplitude/phase modulation, as its modulation scheme. The diagram of the O-QPSK modulator is illustrated in Figure~\ref{fig:ZigBeeMod}. The modulator input comes from a $4$-QAM constellation, where the symbols are $\{\pm{1}\pm{1j}\}$. The real and imaginary parts of the input symbols are processed separately to generate I/Q signals. The quadrature branch of signal samples is shifted by a delay to introduce the offset. As evident in the output waveform, the quadrature branch exhibits a slight lag.

To construct an O-QPSK modulator for ZigBee protocol, we combine the NN-defined QPSK modulator with a shifting process to form the NN-defined O-QPSK modulator, as depicted in Figure~\ref{fig:ZigBeeMod}. We generate symbols from messages following the specification and feed them into our NN-defined O-QPSK modulator. The modulated signals are sent over the air utilizing the prototype in Figure~\ref{fig:TestbedOverview}. We employ the TI CC2650 Kit~\cite{TIpad} as the ZigBee receiver, which can parse the captured signals into messages. 

\begin{figure}[!htbp]
    \vspace{-2mm}
    \centering
    \begin{subfigure}[b]{0.45\linewidth}
        \includegraphics[width=0.8\linewidth, keepaspectratio=true]{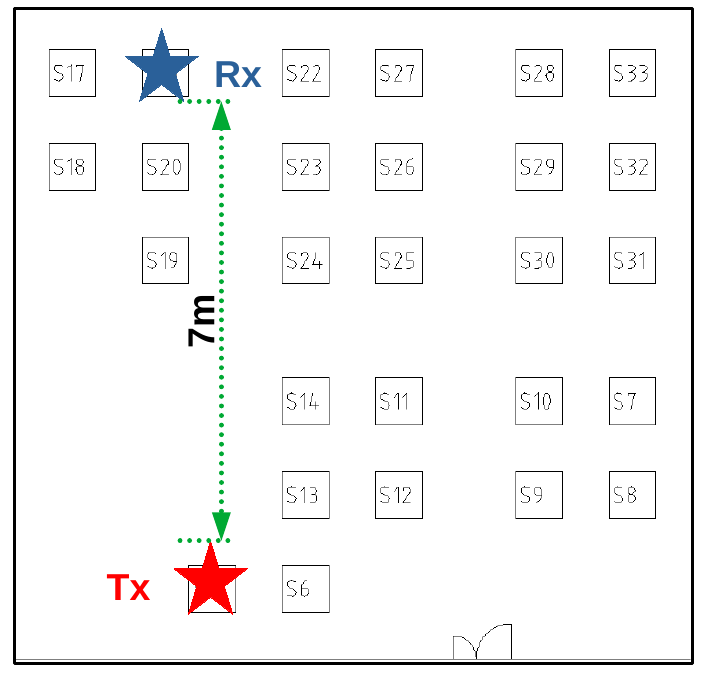}
        % \caption{Porting a QAM-4 modulator deveolped on PC to an ARM-based system}
        \caption{}
        \label{fig:ZigBeeIndoor}
    \end{subfigure}
    %\hfill
    \begin{subfigure}[b]{0.45\linewidth}
        \includegraphics[width=1.1\linewidth, keepaspectratio=true]{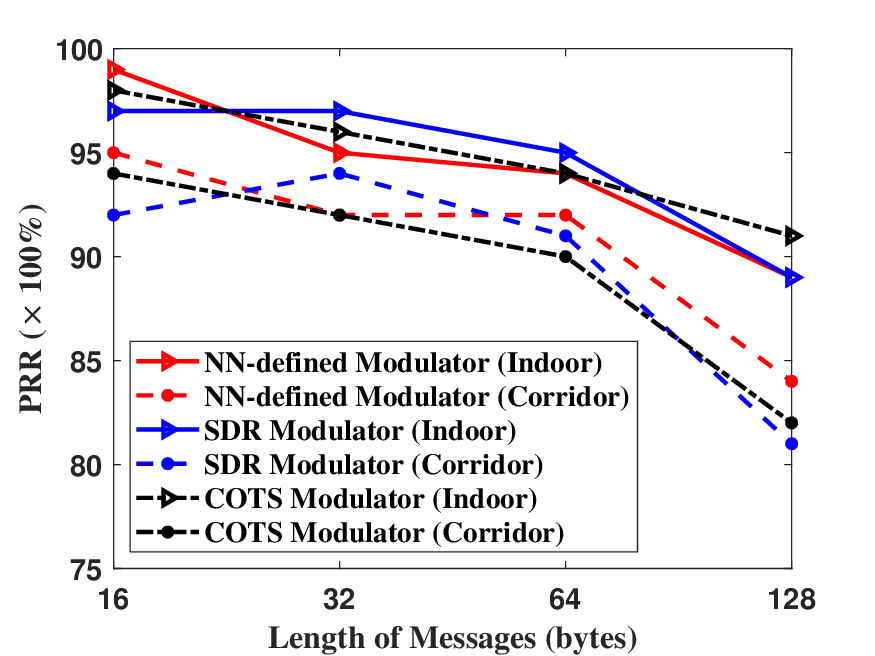}
        % \caption{Porting a QAM-4 modulator deveolped on PC to an ARM-based system}
        \caption{}
        \label{fig:PRRZigBee}
    \end{subfigure}
    \vspace{-2mm}
    \caption{(a) Evaluation settings in indoor environment. (b) Packet Reception Ratio of ZigBee packets modulated by NN-defined modulator and SDR modulator.}
    \vspace{-4mm}
\end{figure}

We generate ZigBee packets with varying message lengths and transmit $100$ packets. At the receiver side, the received ZigBee packets without errors are recorded, and we calculate the packet reception ratio (PRR) in different settings, repeating the evaluation $5$ times. We compare the performance with the SDR implementation using signal processing libraries. And we conduct the same experiments on commercial off-the-shelf~(COTS) TI devices as a baseline. The evaluation is conducted indoors and outdoors. The settings of indoor environments are demonstrated in Figure~\ref{fig:ZigBeeIndoor}. As depicted in Figure~\ref{fig:PRRZigBee}, the ZigBee signals generated by the NN-defined modulator can be successfully received by the commercial device, achieving performance comparable to the existing SDR implementation and commercial devices.

\subsubsection{WiFi-compliant Signals} 
WiFi, which is also extensively utilized for IoT communication, typically employs the OFDM scheme. When implementing the NN-defined modulator for WiFi communication, the process becomes slightly more complex, as WiFi utilizes the CP-OFDM~\cite{Proakis2007digital} modulator, and WiFi frames generally consist of signals generated from various fields.

\begin{figure}[htbp]
    \vspace{-2mm} 
    \centerline{\includegraphics[width=0.7\linewidth, keepaspectratio=true]{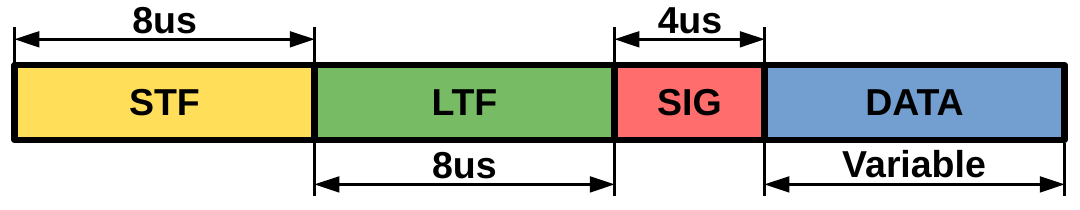}}
    \caption{Fields in IEEE 802.11a/g frame.}
    \label{fig:WiFiField}
    \vspace{-5mm} 
\end{figure}

Taking IEEE 802.11a/g as an example in Figure~\ref{fig:WiFiField}, WiFi frames comprise four fields: Short Training Field (STF), Long Training Field (LTF), Signaling Field (SIG), and Data Field (DATA). The STF and LTF primarily serve detection, synchronization, and channel estimation purposes at the receiver. The SIG contains information about the current frame, such as frame length and modulation and coding scheme information, while the DATA field carries the data.
 % as illustrated in Figure~\ref{fig:WiFiField}

Different fields need specific operations. The STF and LTF involve repeating the signals from the OFDM modulator, while the SIG and DATA require adding a cyclic prefix to the modulated signals by copying the ending parts of the OFDM signals to the front. Following the discussion in Section~\ref{sec:modulator}, we combine an NN-defined OFDM modulator with additional operations to add the cyclic prefix. 
% Figure~\ref{fig:CP_OFDMMod} presents a diagram of an NN-defined modulator for the DATA fields. 
% Additionally, we visualize signal samples generated from the modulator in Figure~\ref{fig:CP_OFDMSignal}. We utilize the $64$-S.C. OFDM modulator, and the ending $16$ points of signal samples are used as the cyclic-prefix. As a result, a total of $80$ signal samples are generated by one OFDM symbol.
% \begin{figure}[htbp]
%     \vspace{-2mm} 
%     \centerline{\includegraphics[width=0.7\linewidth, keepaspectratio=true]{Figures/Evaluation2/OTA_IoT/CP_OFDMMod.pdf}}
%     \caption{Diagram of NN-defined modulator for DATA field.}
%     \label{fig:CP_OFDMMod}
%     \vspace{-4mm} 
% \end{figure}

% \begin{figure}[!h]
%     \vspace{-2mm}
%     \centering
%     \begin{subfigure}[b]{\linewidth}
%         \centerline{\includegraphics[width=0.7\linewidth, keepaspectratio=true]{Figures/Evaluation2/OTA_IoT/CP_OFDMMod.pdf}}
%         \caption{}
%         \label{fig:CP_OFDMMod}
%     \end{subfigure}
%     \vspace{-2mm}
%     %\hfill
%     \begin{subfigure}[b]{\linewidth}
%         \centerline{\includegraphics[width=0.75\linewidth, keepaspectratio=true]{Figures/Evaluation2/OTA_IoT/CP_OFDMSig.eps}}
%         \caption{}
%         \label{fig:CP_OFDMSignal}
%     \end{subfigure}
%     \caption{(a) Diagram of NN-defined modulator for DATA field. (b) In-phase signals from the DATA field modulator.}
%     \vspace{-2mm}
% \end{figure}

\begin{figure}[ht]
    \vspace{-2mm} 
    \centerline{\includegraphics[width=0.8\linewidth, keepaspectratio=true]{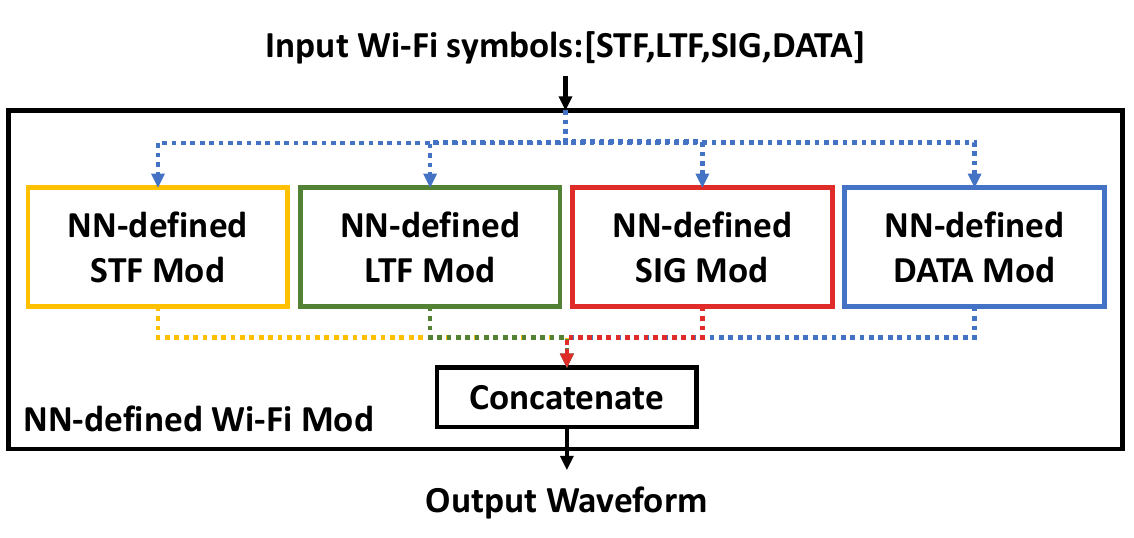}}
    \vspace{-2mm} 
    \caption{NN-defined Wi-Fi modulator.}
    \label{fig:WiFimod}
    \vspace{-5mm} 
\end{figure}

Four NN-defined modulators corresponding to the four fields in IEEE 802.11a/g WiFi frames are implemented. These modulators are then combined to create a single NN-defined WiFi modulator. The overall structure is illustrated in Figure~\ref{fig:WiFimod}. The NN-defined modulators for \texttt{STF}, \texttt{LTF}, \texttt{SIG}, and \texttt{DATA} fields collectively form the NN-defined WiFi modulator, allowing for a comprehensive modulation process that addresses the unique requirements of each field. 

\begin{figure}[!htbp]
    \centerline{\includegraphics[width=0.8\linewidth, keepaspectratio=true]{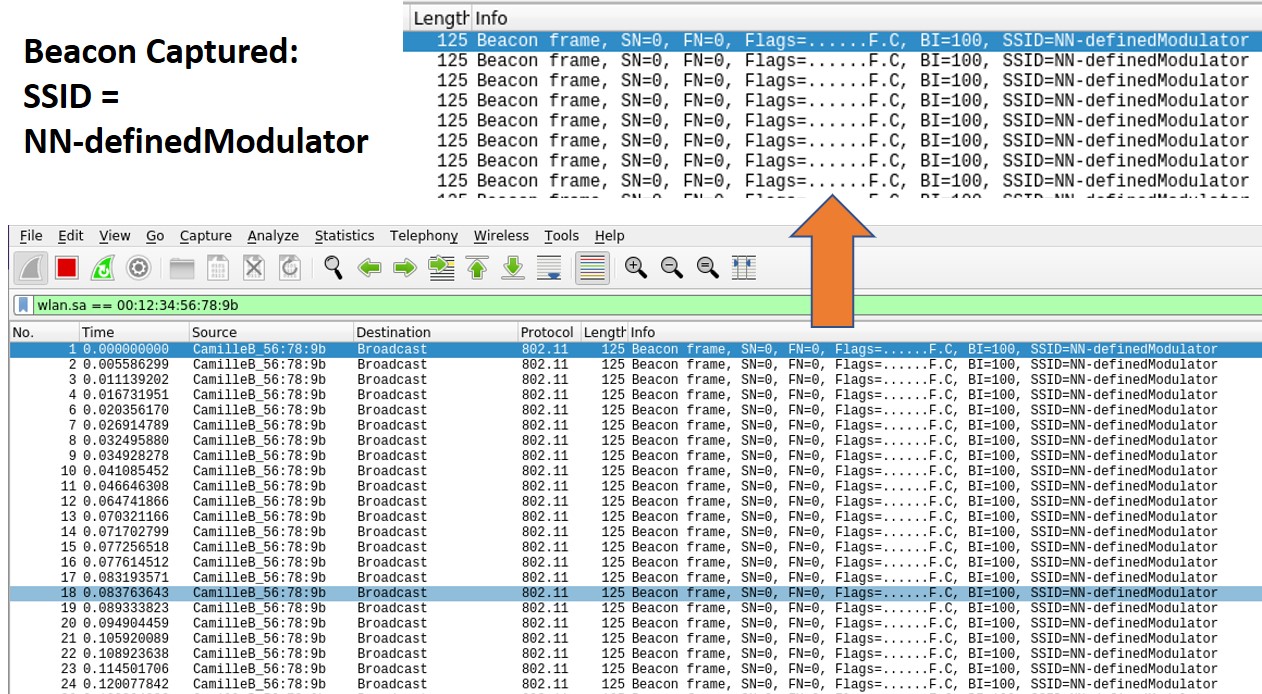}}
    \caption{Reception of beacon signals generated by NN-defined Wi-Fi modulator.}
    \label{fig:WiFiBeacon}
    \vspace{-5mm} 
\end{figure}

We generate beacon packet signals using the NN-defined WiFi modulator and transmit them over the air. A laptop is used to sniff the beacon packets. We test beacon reception in an indoor environment at the $5$GHz band, transmitting $100$ beacon packets for $5$ times. Figure~\ref{fig:WiFiBeacon} demonstrates that the laptop can successfully receive the beacon with an SSID of "NN-definedModulator", achieving a PRR at $96\%$. 

\begin{figure}[!h]
    \vspace{-2mm}
    \centering
    \begin{subfigure}[b]{0.32\linewidth}
        \centerline{\includegraphics[width=0.7\linewidth, keepaspectratio=true]{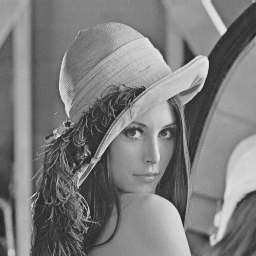}}
        \caption{}
    \end{subfigure}
    \begin{subfigure}[b]{0.32\linewidth}
        \centerline{\includegraphics[width=0.7\linewidth, keepaspectratio=true]{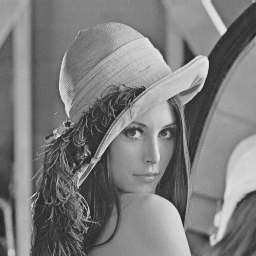}}
        \caption{}
    \end{subfigure}
    \begin{subfigure}[b]{0.32\linewidth}
        \centerline{\includegraphics[width=0.7\linewidth, keepaspectratio=true]{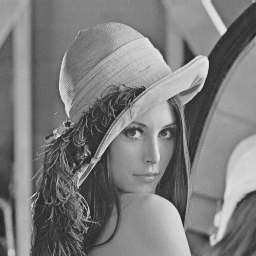}}
        \caption{}
    \end{subfigure}
    \vspace{-1mm}
    \caption{(a) Original image of $256\times256$ pixels; Received images using (b) $16$-QAM at SNR=10dB and (c) using $64$-QAM at SNR=20dB.}
    \vspace{-5mm}
    \label{fig:WiFilena}
\end{figure}

Next, we extend our design to transmit data by generating data packets and passing the signals through simulated AWGN channels. We follow the standard process to detect and synchronize WiFi frames using STF signals, conduct channel estimation and equalization using LTF signals, and then demodulate and decode the SIG and DATA signals. In our evaluation, the symbols for a grayscale image data are generated using $16$-QAM and $64$-QAM. The results are listed in Figure~\ref{fig:WiFilena}. As shown in the figure, we can successfully reconstruct the transmitted images under different settings, further demonstrating the effectiveness and versatility of our NN-defined modulator design in practical applications.

% In summary, this section demonstrates the flexibility and adaptability of NN-defined modulators in addressing practical IoT communication protocols, such as ZigBee and WiFi, by combine the NN-defined modulators with specific operations needed for protocols.

% \begin{figure}[htbp]
%     \vspace{-2mm}
%     \centering
%     \subfigure[a]{0.5\linewidth}{
%         \includegraphics[width=0.8\linewidth, keepaspectratio=true]{Figures/Evaluation2/OTA_IoT/CP_OFDMMod.pdf}
%     }
%     \label{MCS=3, SNR=10dB}
% \end{figure}
    % \quad
    % \subfigure[a]{0.5\linewidth}{
    % \includegraphics[0.5\linewidth, keepaspectratio=true]{Figures/Evaluation2/OTA_IoT/WiFiLena/lena_MCS3_SNR20.png}
    % \label{MCS=3, SNR=20dB}
    % }
    % \quad
    % \subfigure[a]{0.5\linewidth}{
    % \includegraphics[0.5\linewidth, keepaspectratio=true]{Figures/Evaluation2/OTA_IoT/WiFiLena/lena_MCS6_SNR10.png}
    % \label{MCS=3, SNR=20dB}
    % }
    % \quad
    % \subfigure[a]{0.5\linewidth}{
    % \includegraphics[0.5\linewidth, keepaspectratio=true]{Figures/Evaluation2/OTA_IoT/WiFiLena/lena_MCS6_SNR20.png}
    % \label{MCS=3, SNR=20dB}
    % }
    % \caption{Reconstructed images from noisy received signal. MCS-3 applies $16$-QAM constellation and MCS-6 applies $64$-QAM constellation to generate symbols.}
    % \label{fig:WiFilena}

\section{Related Works}\label{sec:rela}

\textbf{SDR solutions for IoT gateway: } SDRs are proposed as universal gateways operating across technologies. Past work \cite{zhang2011wireless, surligas2015empowering, bernardos2014architecture} has developed smart home gateways using the USRP radio with GNUradio support. \cite{narayanan2018revisiting} revisited the SDR-based IoT gateway for decoding collapsed packets. Other solutions employed the cross-technology communication technique as an alternate for IoT gateways~\cite{ wang2020x,jiang2017transparent, li2017webee, jiang2017bluebee,jiang2018achieving, wang2022x, wang2018symbol, wang2019networking,liu2019lte2b,liu2020xfi,liu2021wibeacon}.
% \textbf{Other solutions for IoT gateway: } Cross-technology 

\textbf{Machine learning for communication system: } Neural networks or machine learning has been extensively used in physical layer designs \cite{o2017introduction, soltani2018autoencoder, wen2018deep, gao2018comnet, mu2019end,he2019model,zhu2019joint}. \cite{Data:Oshea:17} introduced a method to learn an end-to-end communication system by interpreting it as an autoencoder \cite{DL:Yoshua}. In \cite{zhao2021deep}, a DNN model replaces all blocks in the conventional OFDM receiver. In \cite{soltani2022OFDMIoT}, the researchers propose to replace processing blocks in the OFDM receiver with neural network models and deploy them on IoT devices.

Our work has innovations in two tiers, distinctive objectives, and different methodologies. Objective-wise, most literature views the neural network as an optimizer and seeks performance gains under complex conditions \cite{zhu2019joint}. In contrast, we use the neural network as an abstraction layer for the portability of IoT gateway functionalities. Methodology-wise, the literature commonly adopts data-driven approaches that employ general-purpose neural networks \cite{wen2018deep, gao2018comnet, soltani2018autoencoder}. We adopt a model-driven approach that designs the neural network-based modulators with reference to the mathematical models. 

\section{Discussion}\label{sec:disc}
It's worth pointing out that we only discuss the linear amplitude/phase modulation schemes in this paper. Other modulation schemes require further study, such as frequency modulation, also known as non-linear modulation. Following the similar idea, We can model the frequency modulation based on the phase changes and construct another NN-defined modulator template that can be used for the Gaussian frequency shift keying (GFSK) modulators used in Bluetooth \cite{heydon2012bluetooth}. Moreover, we intend to extend the application of the learning ability.We can further apply the neural network to learn to reduce the adjacent channel leakage ratio~(ACLR) for single carrier scheme or to reduce the peak-average power ratio~(PAPR) for OFDM scheme. We can also apply the NN-defined modulator to learn from noisy signal samples to reconstruct noiseless modulators. The model-driven approach can also be applied to the receiver design, including demodulation and decoding, which is an emerging topic in wireless communication.

\section{Conclusion}\label{sec:concl}
In this paper, we present an NN-defined modulator template for various modulation schemes that can be converted to a unified NN framework for portable deployment for IoT gateway design. The proposed NN-defined modulator has the extensibility to achieve various modulation schemes for IoT connections, and the evaluation results show that they can perform well. The NN-defined modulator outperforms the existing SDR solutions in terms of portability and efficiency thanks to the wide support of the NN on heterogeneous computing platforms. Meanwhile, the NN-based implementation also enables our design with the learning ability, featuring the potential for intelligent communication systems.
% \input{Sections/TodoList}

%-------------------------------------------------------------------------------

% Lists are sometimes quite handy. If you want to itemize things, feel
% free:

% \begin{description}
  
% \item[fread] a function that reads from a \texttt{stream} into the
%   array \texttt{ptr} at most \texttt{nobj} objects of size
%   \texttt{size}, returning returns the number of objects read.

% \item[Fred] a person's name, e.g., there once was a dude named Fred
%   who separated usenix.sty from this file to allow for easy
%   inclusion.
% \end{description}

% \noindent
% The noindent at the start of this paragraph in its tex version makes
% it clear that it's a continuation of the preceding paragraph, as
% opposed to a new paragraph in its own right.

%-------------------------------------------------------------------------------
\section*{Acknowledgments}
We extend our heartfelt thanks to the reviewers for their rigorous and constructive feedback and the shepherding committee for their invaluable guidance. Special appreciation goes to our collaborator, Shuai Wang from George Mason University, whose diverse skills also significantly enriched this work. This research is supported by the Ministry of Education, Singapore, under its Academic Research Fund Tier 2 (MOE-T2EP20221-0017), National Natural Science Foundation of China under Grant No.62272098 and The Future Network Scientific Research Fund Project (Grant No. FNSRFP-2021-YB-17). This research is also supported by the National Research Foundation, Singapore, and Infocomm Media Development Authority under its Future Communications Research \& Development Programme.
%-------------------------------------------------------------------------------

% %-------------------------------------------------------------------------------
% \section*{Availability}
% %-------------------------------------------------------------------------------

%-------------------------------------------------------------------------------
\bibliographystyle{plain}
\bibliography{Sections/reference}

\end{document}